\documentclass[12pt,preprint]{aastex}
\usepackage{natbib}
\slugcomment{2012 ApJ 751 22}

\begin{document}

\title{A Significant Population of Candidate New Members of the $\rho$ Ophiuchi Cluster}

\author{Mary Barsony\altaffilmark{1,2}}
\affil{SETI Institute, 189 Bernardo Ave, Suite 100, Mt. View, CA 94043, mbarsony@seti.org}

\and

\author{Karl E. Haisch Jr.\altaffilmark{1}}
\affil{Utah Valley University, Physics Dept., 800 W. University Pkwy., Orem, Utah  85058-5999, Karl.Haisch@uvu.edu}

\and

\author{Kenneth A. Marsh}
\affil{Infrared Processing and Analysis Center, California Institute of Technology 100-22, Pasadena, California 91125, kam@ipac.caltech.edu}

\and

\author{Chris McCarthy}
\affil{San Francisco State University, Dept. of Physics \& Astronomy, 1600 Holloway Ave., San Francisco, California 94132, exoplanet@gmail.com}

\altaffiltext{1}{Visiting Astronomer at the Anglo-Australian Telescope, Siding Spring, Australia.}

\altaffiltext{2}{San Francisco State University, Dept. of Physics \& Astronomy, 1600 Holloway Ave., San Francisco, California 94132}

\begin{abstract}
We present a general method for identifying the pre-main-sequence population
of any star-forming region, unbiased with respect to the presence or absence of disks,
in contrast to samples selected primarily via their mid-infrared emission from {\it Spitzer} surveys.
We have applied this technique to a new, deep, wide-field, near-infrared imaging survey of the $\rho$ Ophiuchi cloud core 
to search for candidate low mass members.  In conjunction with published {\it Spitzer} IRAC photometry, and least squares fits of model spectra (COND, DUSTY, NextGen, and blackbody) to the observed spectral energy distributions, we have identified 948
candidate cloud members within our 90\% completeness limits of $J=20.0$, $H=20.0$, and $K_S=18.50$. 
This population represents a factor of $\sim$3 increase in the number of known young stellar objects in the $\rho$ Ophiuchi cloud. 
A large fraction of the candidate cluster members (81\% $\pm$ 3\%) exhibit infrared excess emission consistent with the presence of disks, thus strengthening the possibility of their being {\it bona fide} cloud members. Spectroscopic follow-up will confirm the nature of individual objects, better constrain their parameters, and allow an initial mass function to be derived.
\end{abstract}


\section{Introduction}
Sub-stellar objects, including brown dwarfs, sub-brown dwarfs, and free-floating
objects of planetary mass, are all at their most luminous (by orders of magnitude) upon formation.
Therefore, the nearest and youngest star-forming regions (SFRs) present the best 
opportunity to determine the shape of the initial mass function (IMF) at the lowest
masses.  Theories of star-formation that attempt to predict IMFs 
\citep{KRO11, HEN11, MCK07}, require observational constraints, especially at the low mass end.
Specifically, one pressing open question is:  What is the lowest mass object that can form via the usual star-formation process?

The advent of large-format near-infrared (NIR) array detectors has opened the door
to surveying large angular extents of nearby SFRs with 
sensitivity sufficient to uncover planetary mass objects.  The potential now exists
to determine the IMF shapes of the nearest and youngest SFRs through the substellar regime down to
$\sim$1$-$2 M$_{Jup}$. 

Recent work on some nearby SFRs has determined IMFs in a statistical
sense only, by deriving luminosity functions (LFs), deduced from imaging data
obtained in just a single NIR waveband.  
For example, IMFs were derived from observed K-band luminosity functions (KLFs)
for the Orion Nebula Cluster \citep[ONC--][]{MUE02}
and IC348 \citep{MUE03} or from the observed J-band luminosity function (JLF) 
for IC348 \citep{PRE03}.  This method for IMF determination relies on: 
$i)$ an assumed form of the mass function, 
$ii)$ an assumed age distribution, and $iii)$ pre-main-sequence
model tracks which give the single filter brightness for a given mass. Additional complications that arise for this (and other) method(s)
of IMF determination include: distinguishing 
cluster members from non-members, accounting for the extinction to each object, 
and correcting for excess emission due to circumstellar material/disks to the observed KLF's (this latter effect is minimized for JLFs). 

A summary of published IMFs that make use of NIR photometric data for six nearby SFRs is presented in Table 1.
Distances, ages, telescope/instrument combinations used, completeness limits reached, and references are listed.
The potential for deriving an IMF in $\rho$ Oph is shown by listing the parameters of our survey in the last entry of Table 1.

Alternative approaches for IMF determinations of SFRs, applicable to low extinction regions, include converting the observed JLF to an LF with the aid of model pre-main-sequence
tracks by applying bolometric corrections (B.C.'s) derived from field stars.
Pre-main-sequence tracks, for a presumed cluster age, are then
used to convert the luminosity function to a mass function 
\citep[e.g.,][]{LOD09} for $\sigma$ Ori.
This approach is aided by complementary spectroscopy,
since the B.C.'s are a function of spectral type \citep[e.g.,][]{CAB07}.
Optical multi-object spectroscopy has been used for IMF determinations
in relatively low-extinction regions of SFRs  to
derive cluster membership, extinction, and spectral types for individual candidate 
young stellar objects (YSOs). Broadband photometry is then used, in conjunction with
bolometric corrections, to derive individual YSO luminosities. Each YSO is plotted on an
H-R diagram, and comparison with a set of pre-main-sequence tracks is used to derive a mass and age for
each cluster member \citep[e.g.,][]{LUH04} for Taurus; \citep[][]{LUH07} for Cha I. 

Finally, fitting of multi-wavelength spectral energy distributions (SEDs) has been used to derive de-reddened effective 
temperatures (T$_{eff}$) for cloud members. Pre-main-sequence tracks, for an assumed cluster age, are then used
to derive mass values for a given T$_{eff}$ \citep[e.g.,][]{MAA10} for $\rho$ Oph.
This approach, however, is problematic for treating YSOs with circumstellar disks and/or envelopes.

Related to the determination of the shape of the IMF at the lowest masses
is the search for a low mass cut-off of the IMF.  This search has been
the focus of many recent investigations.  We list those
making use of deep, NIR imaging in Table 2.  In the IC348 cluster, 
one new candidate planetary mass object (PMO) has been proposed but awaits
spectroscopic confirmation \citep{BUR09}.  
The most recent deep, large-area, NIR study of the ONC,
covered a 30$^{\prime}\times40^{\prime}$ region to $J\simeq$ 19.5 mag, $H\simeq 18.0$ mag, and $K_s\simeq 18.5$ mag
(3$\sigma$), a sensitivity sufficient to detect 1 MYr old PMO's to $A_V\simeq 10$ mag of extinction \citep{ROB10}.
These authors found 1298 
sources in the reddened
brown dwarf region and 142 
in the reddened PMO region of the $H$ vs. $J-H$  
color-magnitude diagram.   The source distribution in the 
corresponding regions of the $K_s$ vs. $H-K_s$
color-magnitude diagram yielded 2134
sources in the reddened
brown dwarf region and 421 sources in the reddened PMO region. 
In NGC1333, no PMO's have been found, but 19
spectroscopically confirmed brown dwarfs have, from a sample of 36 objects with $i'-z'$ colors expected for
young, very low mass objects \citep{SCH09}.  This, despite the sensitivity of the survey
to mass limits of 0.008 M$_{\odot}$ for $A_V\le 10$ mag and 0.00 4M$_{\odot}$ for $A_V\le5$ mag, led the authors to the conjecture
that the low-mass cut-off corresponding to T$_{eff}<$ 2500 K
has been found for this cluster. 
In the ChaI SFR, a deep optical imaging survey, with follow-up deep 
near-infrared photometry, sensitive to mass limits of 0.003 M$_{\odot}$ $-$0.005 M$_{\odot}$ for $A_v \le 5$, with follow-up 
low-resolution optical spectroscopy \citep{MUZ11}, found no new confirmed substellar objects, beyond those found in \citet{LUH07}.  The more recent study
placed upper limits on the number of missing planetary mass members  (down to $\sim$ 0.008 M$_{\odot}$)
of $\le$ 3\% ($\le$ 7) of the currently known YSO population in Chamaeleon I.
Due to its large angular extent, searches for PMO's in the Taurus SFR have been targeted towards known members, be they young stellar objects (YSOs) or young brown dwarfs. Two notable PMO candidates discovered in this way, necessarily members of multiple systems, are TMR1-c 
\citep{TER98, RIA11} and 2MASS J04414489$+$2301513 \citep{TOD10}.
Large area surveys of Taurus using 2MASS and {\it Spitzer} data have been used to search for new members, including brown dwarfs
\citep{LUH06, LUA09, LUB09, REB10}, bringing the total list of known Taurus members to 318 \citep{LUH10},
of which 43 are spectroscopically confirmed as substellar \citep{MON10}.
In the $\sigma$ Ori cluster, a careful re-analysis of archival data, supplemented by newer large-area, sensitive imaging,
led to the discovery of three new planetary mass candidates
down to $\sim$ 4 $M_{Jup}$, bringing to 17 the total number of candidate PMO's in this region \citep{BIH09}. 
In the recent deep NIR imaging study of Lupus III, no planetary mass objects (corresponding to late T spectral types),
but 17 sub-stellar candidates (with 1700K $\le$ T$_{eff}$ $\le$ 3000K)
have been identified \citep{COM11}.

The subject of this study, the $\rho$ Oph SFR, is an especially
attractive target, due to its relatively compact area,
proximity (d$=$120 $\pm$ 5 pc--\citet{LOI08}, 
richness of its embedded young
stellar population \citep{WIL08, BAR05}, and youth $\sim$ 1 MYr
\citep{LUH99, PRA03, WIL05}.
As such, it has been the target of several, deep, large area NIR surveys whose stated aim
is finding its lowest mass members.  In their sensitive NIR survey of $\sim$1 deg$^2$ of the main $\rho$ Ophiuchi cloud core, \citet{ALV10} report the detection
of $\sim$ 5.7 $\times$10$^4$ objects, which they winnowed down using various color-magnitude diagrams
guided by the 1 MYr DUSTY isochrone of \citet{CHA00}. Their final list of 110 candidate substellar members
includes 80 which are newly identified. 
Another recent survey searching for the lowest mass member in $\rho$ Oph
covered a 31.5$^{\prime} \times 26^{\prime}$ area in the $iJK_s$ filters. This investigation resulted 
in the discovery of one new, spectroscopically confirmed, brown dwarf 
\citep{GEE11}.  Furthermore, these same authors identified
27 brown dwarf candidates (11 of which have previous spectroscopic confirmation) using {\it Spitzer} photometry.
Analysis of the 2MASS calibration strip, running along a 9$^{\prime}$ wide swath through the $\rho$ Oph core, resulted in the identification
of 11 possible planetary mass objects (T $\le$ 1800K) of $\sim$ 115 cluster-member candidates (Marsh et al. 2010a). Follow-up spectroscopy
of seven of these resulted in the discovery of a single planetary mass object 
\citep{MAB10}.

An alternate approach that has been used to search for the lowest mass member of the $\rho$ Oph core is 
to target the spectroscopic signature of methane (CH$_4$) absorption, found in low-temperature (T$_{eff}\le$1500 K)
atmospheres.  Images are acquired through specially designed adjacent narrow-band 
methane-off ($\sim$1.6$\mu$m) and methane-on ($\sim$1.7$\mu$m)
filters covering the $H$-band. Methane absorbing objects, that would have masses of $\sim$ 1$-$2 M$_{Jup}$
at the age and distance of the $\rho$ Oph core, would appear uniquely and characteristically bright in methane-off minus
methane-on differential images.  This technique has been used to survey
a 920 arcmin$^2$ region of the $\rho$ Oph core to identify 22 planetary mass candidates \citep{HAI10}.

Spectroscopic follow-up of these candidates is currently being carried out by our team. As a part of our ongoing
$H-$band methane-filter imaging program of nearby SFRs to discover their lowest mass 
members, we are also acquiring complementary deep $J$ and $K_s$ data. 
In this paper, we report the discovery of 948 candidate low-mass members from combined, deep, $JHK_s$ imaging
of the central 920 arcmin$^2$ area of the $\rho$ Ophiuchi  cloud core, supplemented by {\it Spitzer} data.

The new observations and data reduction methods are described in $\S$2. Results using modelling of our 
deep $JHK_s$ photometry combined with {\it Spitzer} photometry are described in $\S$3. Properties of 
the newly discovered sources, estimation of sample contamination, and detailed comparisons with recently
published deep, large-area NIR imaging surveys are contained in $\S$4.  The summary and conclusions are presented 
in $\S$5.

\section{Observations and Data Reduction}
Observations of the $\rho$ Ophiuchi cloud core were obtained during the period 2008 May 23 - 26 with the IRIS2 NIR imager/spectrograph on the Australian Astronomical Observatory's
4 m telescope (AAT). IRIS2 consists of a Hawaii HgCdTe 1024$\times$1024 array which, when mounted at the f/8 Cassegrain focus on the AAT, yields a plate scale of 0\farcs45 pixel$^{-1}$ with a corresponding field of view of approximately 7.7$^\prime$$\times$7.7$^\prime$. For all observations, the $J, K_{s}$ (1.25, 2.14 $\mu$m), CH$_{4s}$ (1.59 $\mu$m), and CH$_{4l}$ (1.673 $\mu$m) filters were used. Details of our observation and data reduction procedures for all filters are discussed in \citet{HAI10}. However, because this paper makes extensive use of the $J$ and $K_{s}$ data, which the previous work did not, we summarize below the observations and image reduction process for these filters.

Nineteen IRIS2 fields, centered at $\alpha$ = 16$^{h}$26$^{m}$56.34$^{s}$, $\delta$ = -24$^{\circ}$28$^\prime$52\farcs22 (J2000), were observed in a rectangular pattern covering an area of $\sim$ 920 arcmin$^{2}$ on the sky. The observed
area is shown by the solid outlines in Figure 1, superposed on the extinction map that was derived from the 
2MASS catalog\footnote[3]{http://www.cfa.harvard.edu/COMPLETE/data\underline{\ \  }html\underline{\ \  }pages/OphA\underline{\ \  }Extn2MASS\underline{\ \  }F.html}
as part of the COMPLETE project \citep{RID06, LOM08} using the NICER algorithm \citep{LOM01}.
Each field was spatially overlapped by 1 arcminute in both right ascension and declination to allow for redundancy of photometric measurements of sources located in the overlapped regions. All fields were observed in the Mauna Kea Observatory (MKO) photometric system $J$ and $K_{s}$ filters in a five point dither pattern with 30\arcsec \hspace*{0.05in}offsets between each dither. Integration times at each dither position for the $J$ and $K_{s}$ filters were 15 seconds $\times$ 4 coadds and 6 seconds $\times$ 10 coadds, respectively, for a total integration time of 5 minutes in each filter. The FWHM for all observations varied between approximately 2.2 - 3.1 pixels ($\sim$ 1\farcs0 - 1\farcs4), with the worst seeing being at the shortest wavelengths ({\it e.g}., the $J$ filter).

All data were reduced using the Image Reduction and Analysis Facility (IRAF)\footnote[4]{IRAF is distributed by the National Optical Astronomy Observatories, which are operated by the Association of Universities for Research in Astronomy, Inc., under cooperative agreement with the National Science Foundation.}. An average dark frame was constructed from the dark frames taken at the beginning and end of each night's observations. This dark frame was subtracted from all target observations to yield dark subtracted images. Sky frames in each filter were individually made for each observation by median-combining all five $J$ and $K_{s}$ band frames for each field. The individual sky frames were normalized to produce flat fields for each target frame. All target frames were processed by subtracting the appropriate sky frames and dividing by the flat fields. Target frames were then registered and combined to produce the final reduced images in each filter. $H$-band images of each field were constructed by adding the corresponding CH$_{4s}$ and CH$_{4l}$ images for the particular field in question.

Infrared sources were identified at $K_{s}$-band using the automated source extractor DAOFIND routine within IRAF \citep{STE87}. DAOFIND was run on each field using a FWHM of 2.8 pixels, and a single pixel finding threshold equal to 3 times the mean noise of each image. Each field was individually inspected, and the DAOFIND coordinate files were edited to remove bad pixels and any objects misidentified as stars, as well as to add any missed stars to the list. Objects within 30\arcsec \hspace*{0.05in}of the field edges were also removed from the list, as they were in low signal to noise regions of the image as a result of the dither pattern used. Aperture photometry was then performed on all fields in each filter using the PHOT routine within IRAF. An aperture of 4 pixels in radius was used for all target photometry, and a 10 pixel radius was used for the standard star photometry. Sky values around each source were determined from the mode of intensities in an annulus with inner and outer radii of 10 and 20 pixels, respectively. Our choice of aperture size for our target photometry insured that the individual source fluxes were not contaminated by the flux from neighboring stars; however, they are not large enough to include all the flux from a given source. In order to account for this missing flux, aperture corrections were determined using the MKAPFILE routine within IRAF. The instrumental magnitudes for all sources were corrected to account for the missing flux.

Photometric calibration was accomplished using the list of standard stars of \citep{PER98}. The standards were observed on the same nights and through the same range of air masses as the $\rho$ Ophiuchi cloud. Zero points and extinction coefficients were established for each night.\footnote[5]{We found a zero point offset of 1.8 mag through each filter.} All magnitudes and colors were transformed to the CIT system using MKO to 2MASS and 2MASS to CIT photometric color transformation equations\footnote[6]{See http://www.astro.caltech.edu/$\sim$jmc/2mass/v3/transformations/}, and the conversion relations of \citet{STE04}. Because of the extensive spatial overlapping of the cloud images, a number of sources were observed at least twice. We compared the $JHK_{s}$, magnitudes of 200 duplicate stars identified in the overlap regions. For all stars brighter than the completeness limit of our survey, the photometry of the duplicate stars agreed to within 0.15 magnitudes. Plots of our photometric errors as a function of magnitude are presented in Figure 2, for $J$ (left panel), $H$ (middle panel), and $K_s$ (right panel).  

The completeness limit of our observations was determined by adding artificial stars at random positions to each of the 19 fields in all four filters and counting the number of sources recovered by DAOFIND. Artificial stars were added in twelve separate half-magnitude bins, covering a magnitude range of 16.00 to 22.00, with each bin containing 100 stars. The artificial stars were examined to ensure that they had the same FWHM as the real sources in each image. Aperture photometry was performed on all sources to confirm that the assigned magnitudes of the added sources agreed with those returned by PHOT. All photometry agreed to within 0.10 mag. DAOFIND and PHOT were then run and the number of identified artificial sources within each half-magnitude bin was tallied. This process was repeated 20 times. We estimate that our survey is 90\% complete to $J$ = 20.00, $H$ = 20.00, and $K_{s}$ = 18.50. Furthermore, saturation of objects in each image occurred at J $\simeq$ 12.0, H $\simeq$ 11.0, and K$_{s}$ $\simeq$ 10.0, respectively. Thus our observations are sensitive to 12.0 $\leq J \leq$ 20.0, 11.0 $\leq H \leq$ 20.0, and 10.0 $\leq K_{s} \leq$ 18.5, respectively.

Coordinates for all objects were determined relative to the positions of known objects in the 2MASS\footnote[7]{This publication makes use of data products from the Two-Micron All-Sky Survey, which is a joint project of the University of Massachusetts and the Infrared Processing and Analysis Center/California Institute of Technology, funded by the National Aeronautics and Space Administration and the National Science Foundation.} catalog. In particular, plate solutions were done using the 2MASS catalog in conjunction with WCSTools, a package of programs and a library of utility subroutines for setting and using the world coordinate system in the headers of the most common astronomical image formats to relate image pixels to sky coordinates.\footnote[8]{http://tdc-www.harvard.edu/wcstools/} The resulting coordinates of all objects in our survey have typical rms uncertainties of $\sim$ 0\farcs20 relative to the coordinates of previously known stars used in their determinations.

\section{Results}

We detected a total of 2283 sources at all three wavelengths at or brighter than our $JHK_{s}$ completeness limits within the 
920 arcmin$^2$ region outlined by the solid black lines in Figure 1. 
Of course, many more sources were detected at each individual waveband.
At $K_s$, we detected 7081 sources to 5$\sigma$, and 6882 to our $K_s=18.50$
completeness limit. We used the locations of the $K_s$ detections to search for counterparts
at $H$ and $J$.  Since we used the combination of the 30-min. on-source duration observations in the methane-on and methane-off filters to construct the $H$-band image,  instead of just the 5 minutes total 
on-source integration times at $K_s$ and $J$-bands, we detected 7090 sources at $H$ to 5$\sigma$, and 6986 to our $H=20.0$ completeness limit. Finally, we detected just 3486 sources to 5$\sigma$
at $J$, and 2404 to our $J=20.00$ completeness limit, reflecting the fact that we were observing through the highest extinction portions of the $\rho$ Oph core. 

Our survey boundaries are indicated in Figure 1, superposed on the extinction
map (described in $\S$ 2), displayed in both greyscale and by contour levels.  From Figure 1, it is evident that
our survey encompassed the highest extinction portions of the $\rho$ Oph cloud core.  The vast majority of detected sources lie between the $A_V=5$  and $A_V=15$ contour levels, whereas much of the surveyed area has $A_V\ge20$.

In Figure 3 (left panel), we present the $J-H\ vs.\ H-K_{s}$ color-color diagram for all 2283
objects with 10 $\le K_s \le$ 18.5 in our survey area, with available photometry at all three ($JHK_s$)
bands in our data.
The greatest uncertainty in the colors is less than 0.2 magnitudes for all sources and 
is indicated by the size of the cross in each panel of Figure 3.  The solid curve in each panel
represents the locus of colors corresponding to unreddened main sequence stars, ranging in spectral  type from  early O to M9, after converting the 2MASS colors to the CIT system.
The locus of the colors of giant stars is represented by a dashed line in each panel
\citep{BES88}. 
The two parallel dashed lines define the reddening band for main sequence stars and are parallel to the reddening vector. The classical T Tauri star (CTTS) locus in these diagrams extends from $J - H$ = 0.81, $H - K$ = 0.50 to $J - H$ = 1.10, 
$H - K_{s}$ = 1.00 (Meyer, Calvet, \& Hillenbrand 1997). A diagonal arrow representing the effect of 5 magnitudes of visual extinction is also shown. The reddening law of \citet{COH81}, derived in the CIT system and having a slope of 1.692, has been adopted. 

Note the offset of the detected sources in the left panel of Figure 3 from the (0,0)
position in the color-color diagram, indicating all sources suffer {\it at least} $A_V=$ 5 mag
of visual extinction--confirming that we are looking through the darkest portion
of the $\rho$ Oph cloud core.
Of the 2283 stars plotted in the left panel of Figure 3, 1139 de-redden to the CTTS locus.
Therefore, these objects possess infrared excess emission, and are referred to as ``excess sources'' in the following.
Among the 1139 excess sources, 830 have available {\em Spitzer} photometry.
We have divided the remaining sources into two groups.
The first group consists of 709 ``non-excess'' sources--those which definitely
would not de-redden to the CTTS locus.  The second group, consisting of 435 sources,
would de-redden to photospheric colors characteristic of very low-mass stars or brown dwarfs of 
spectral types in the range M7$-$L0 as given by \citet{LUH10}.
Of the 709 sources which do not display infrared excesses, 533 have available {\em Spitzer} photometry. 
Of the 435 sources which would de-redden to M7$-$L0 photospheric colors,
378 have available {\em Spitzer} photometry.

We have estimated the effective temperatures for the 1723 sources for which successful
fits were obtained to both our $JHK_s$ photometry and to the {\em Spitzer}
mid-infrared photometry.  {\em Spitzer} photometry is taken from
either the c2d (the {\it Spitzer} ``From Molecular Cores to Planet-Forming Disks'' Legacy Program)
CLOUDS catalog for L1688\footnote[9]{http://irsa.ipac.caltech.edu/data/SPITZER/C2D/clouds.html}
or, from \citet{GUT09}, for sources not present in the c2d catalog.

The combined NIR and {\em Spitzer} photometry
for each source was fit with a model spectrum to estimate its effective temperature, $T_{eff}$.
Four possible models were
used for each source.  These were: {\it i)} the 1 MYr COND models for 
$T_{eff} \le 1500K$ \citep{BAR03}, {\it ii)} the 1 MYr DUSTY models for 
$1500K \le T_{eff} \le 3000K$ \citep{CHA00}, {\it iii)} the NextGen
models for $T_{eff}\ge 1700K$, with solar gravity and metallicity 
\citep{HAU99},
or {\it iv)} blackbody spectra for all possible temperatures.
De-reddened $K_s$ magnitudes (for d$=$ 124 pc) were derived using the observed $K_s$ magnitudes
and the $A_V$ estimates obtained by de-reddening each source to either the main-sequence (within
the reddening band of Figure 3) or to the CTTS locus (for sources to the right of the reddening band of Figure 3).
Figure 3 (right panel) shows the distribution of sources in the $JHK_s$ color-color diagram which
are found to lie above the main-sequence from our SED fits.

In Figure 4, we present the dereddened $K_{s}$ magnitude as a function of estimated effective temperature for the 827 ``excess'' (top panel)
and the 527 ``non-excess''  (middle panel) sources for which successful fits were obtained
to their spectral energy distributions determined by our new $JHK_s$ data combined
with {\it Spitzer} data. 
Objects in the cloud-exterior region, from the same off-cloud region as in
\citet{MAA10},
are plotted in the bottom panel of Figure 4, for comparison. 
This bottom panel shows the results of fits to 509 off-cloud sources
for which both deep $JHK_s$ and {\it Spitzer} photometry are available.

All three panels of Figure 4 show model curves for the 1 Myr COND (dashed) and DUSTY (dotted) models, and for main sequence stars (solid) for an assumed distance of 124 pc.  The $T_{eff}$ ranges used for the different atmospheric models are color-coded, and indicated at the bottom of each panel.
Sources best-fit using the 1 MYr COND models are plotted in red, those best-fit
using the 1 MYr DUSTY models are plotted in green, and those best-fit using the NextGen models
are plotted in mustard.  In addition, fits to blackbodies of a specified temperature are plotted in blue.

Details of the fitting procedure are described in 
\citet{MAA10}, with the improvement in the present work
that the {\it Spitzer} filter bandpasses have been convolved with the model atmospheres to derive
{\it Spitzer} IRAC magnitudes for each model.  
For purposes of the model-fitting described above, $A_V$ values were assigned to be those derived from de-reddening each source
to the classical T-Tauri (CTTS) locus of Figure 3, for the ``EXCESS'' sources, and those derived by de-reddening to the main-sequence
for the ``NON-EXCESS'' sources,  instead of letting $A_V$ be a free parameter.

In order to test the validity of our fitting procedure,
we plot the locations assigned by our SED-fitting program 
to known, spectroscopically confirmed, brown dwarfs in $\rho$ Oph from the tabulation of \citet{GEE11}.
In Figure 4, the open diamonds represent known
brown dwarfs with IR excesses (top panel)
and known brown dwarfs without IR excesses (middle panel).
In both cases, our fitting would independently determine these objects to lie above
the main-sequence, and to have low values of $T_{eff}$.

\section{Discussion}
\subsection{Candidate New Members and Their Properties}

The identification of the new candidate cloud members in $\rho$ Oph is primarily based on their location in the plots of Figure 4.
The top panel of Figure 4 shows a dramatically different distribution of sources from those in the middle and bottom panels.  Note the dearth of reddened main-sequence stars in the top panel combined
with the presence of disks, inferred from the 
preponderance of blackbody best-fits. In the top panel of Figure 4, 764
of the 827 successfully fit infrared excess sources lie above the main sequence, identifying them as pre-main-sequence objects, and thus as candidate cluster members.  The open diamonds plotted in the top panel of Figure 4 represent fits to the photometry of previously
known and spectroscopically confirmed brown dwarf members of the cluster that also display infrared excess emission
from disks, for comparison. 

An artifact in the top panel of Figure 4 is the presence of a gap in the distribution of $T_{eff}$ values
from the model fits in the 1200 K $\le T_{eff}\le$ 1800 K range.  The root cause of this gap is that objects surrounded by circumstellar material, and, therefore, exhibiting spectral energy distributions (SEDs) characteristic of disks or late-stage protostars, were fit to purely photospheric models (COND, DUSTY, NextGen, Blackbody).  At the lowest temperatures ($T_{eff} \le$1500 K), the COND models tend to give fairly flat-looking SEDs which often provide artificially good fits to flat spectrum protostars, while the more evolved young stellar objects are better fit by higher-temperature models which are more Planck-like (DUSTY, NextGen, Blackbody).  It can be seen from the top panel of Figure 4 that the distinct gap in the distribution of sources in the 1200 K $\le T_{eff}\le $1800 K range is not an artifact of the 1500 K boundary between the COND and DUSTY models -- rather, this gap represents the temperature range over which none of the photospheric models can adequately mimic circumstellar disks.  

A substantial fraction of the newly discovered population of ``excess'' sources plotted in the top panel of Figure 4 seems concentrated above the lowest $T_{eff}$ NextGen models, but below the 
COND/DUSTY models.  This is very likely due to suppressed K-band flux due to extinction of the YSO photospheres
by cool disk material \citep[e.g.,][]{MAY10}.

In the middle panel of Figure 4, we plot the 527 ``non-excess'' sources for which successful SED fits were obtained
to our combined $JHK_s$ and available {\it Spitzer} photometry.  Among these ``non-excess'' sources, most (343/527)
lie in the region below the main-sequence at the cloud's distance in the de-reddened $K_s$ vs. $T_{eff}$ plot.  Therefore, most of 
the non-excess sources are consistent with being background objects.  However, 184 
of the 527 non-excess sources plotted in the middle panel of Figure 4 lie above the main-sequence,
and are candidate cloud members. The open diamonds plotted in the middle panel of Figure 4
show the locations of fits to the photometry of previously known, spectroscopically confirmed, brown dwarf members of the cluster, that lack infrared excess emission. 

The bottom panel of Figure 4 shows the distribution of sources from an off-cloud region
(same off-cloud region as used by Marsh et al. 2010a, to 5$\sigma$ detection limits of $J=20.5$, $H=20.0$, and $K_s=19.0$).
The majority of objects detected in the off-cloud region lie below the locus of main-sequence photospheres
at the distance to $\rho$ Oph.  Sources falling below the main-sequence locus in Figure 4 are reddened background stars. 

We therefore find a total of 948 candidate young stellar objects (YSOs) in the $\rho$ Ophiuchi cluster, of which 764 are excess sources, and 184 are non-excess sources.  Table 3 lists these sources. Column 1 of Table 3 is an ordinal source identification number,
followed by each candidate object's $\alpha$(2000) and $\delta$(2000) coordinates.  We then list our near-infrared photometry
in the order, $J$, $\sigma_J$, $H$, $\sigma_H$, $K_s$, $\sigma_{K_s}$, followed by the IRAC photometry with corresponding errors in  each of the four IRAC bands (3.6 $\mu$m, 4.5 $\mu$m, 5.8$\mu$m, and 8.0 $\mu$m, respectively) in ascending wavelength order.
The next column lists the extinction values (A$_V$) derived by de-reddening each source 
to either the main-sequence (for ``non-excess'' sources) or to the CTTS locus (for ``excess'' sources)
in the $J-H$ vs. $H-K_s$ color-color diagram of Figure 3.  The next column lists the best-fit value of $T_{eff}$
derived from model fitting to the SED of each source. The penultimate column lists the best-fit model type used to derive the tabulated $T_{eff}$
value. The last column indicates whether an individual source is an ``excess'' source (E) or ``non-excess'' source (NE).

The right panel of Figure 3 shows a plot of the location of these 948 candidate cluster members 
in the $J-H$ vs. $H-K_s$ color-color diagram.   A large fraction, 764/948 or
81\%, of our candidate members lie in the infrared excess region of the $JHK_{s}$ color-color diagram in Figure 3. Predictions from both observations and modeling suggest that this is what one would expect for excess emission from circumstellar disks \citep[e.g., ][]{LAD92, MEY97, HAI00}. If the infrared excesses do indeed originate in circumstellar disks, then this strengthens their identification as a significant population of new low mass YSOs in the $\rho$ Ophiuchi cloud.

Figure 5 shows fit results for the subset of the 435 sources which could be de-reddened to 
very low-mass stellar or brown dwarf colors in the $J-H$ vs. $H-K_s$ color-color diagram of Figure 3,
for which good SED fits could be obtained.
The three panels of Figure 5 illustrate the fact that, for this sample, the fraction of objects
inferred to be pre-main-sequence, and,
therefore, to be potential cluster members, varies greatly with their assumed, intrinsic,
unreddened colors, or, equivalently, with their derived values of $A_v$.
For these
new fits, the {\it a priori} values of extinction were based on de-reddening sources to the photospheric $JHK_s$ colors given for the indicated intrinsic spectral type by Luhman {\it et al.} (2010) in the $J-H$ vs. $H-K_s$ diagram.  {\it Spitzer} photometry was available for 378 of these 435 sources,
and successful fits were obtained for 373 (top panel), 368 (middle panel),
and 372 (bottom panel) sources.  The top panel shows the results of assuming an intrinsic
spectral type of M7 for all fitted sources.  In this case, only 81/373 fitted sources would be
inferred to lie above the main-sequence.  The middle panel shows the results
of assuming an intrinsic spectral type of L0 for all fitted sources.  In this case, 348/368 fitted
sources would be inferred to lie above the main-sequence.  Finally, the bottom panel
shows the results of assuming an intrinsic spectral type somewhere between M7 and L0,
close to M9.  In this case, 312/372 well-fit sources would be inferred to lie above the main-sequence.
Due to this large variation in the inferred fraction of pre-main-sequence sources depending on
the assumed intrinsic spectral type of each source for sources lying in this narrow range of the $J-H$
vs. $H-K_s$ color-color diagram, we do not yet include any of these among the new, low-mass,
YSO population of $\rho$ Oph.  Follow-up spectroscopy will reveal the intrinsic spectral
types of this subset of objects, and will determine what fraction are cloud members.

%
%

For our 948 candidate cloud members, however, we can estimate the range of masses to which this survey is potentially sensitive, given the distribution of $JHK_s$ brightnesses.
At the low mass end, our $JHK_s$ survey is sensitive to a bare photosphere
with T$_{eff}$ $\sim$ 1100 K for the 1 MYr COND models
at the distance to $\rho$ Oph, with no reddening,
corresponding to  $\sim$ 1.5 M$_{Jup}$.  This rises to 2.0, 4.0, and 8.5 M$_{Jup}$ for $A_V=5$, 10, and 15, respectively.
At the high mass end, 
a source with K$_s \sim$14.0 corresponds to an absolute K$_s$
of 8.52 or a T$_{eff}$ $\sim$ 2250 K -- which is a late M spectral type. 
This $K_s$ magnitude corresponds to a mass of 10 M$_{Jup}$ for the 1 MYr COND model, and
rises to 15, 35, and 45 M$_{Jup}$ for $A_V=5$, 10, and 15, respectively.  However, these estimates do not take the complicating factor of disk
excesses into account. Spectroscopic follow-up is required to confirm the nature of individual objects, and to better constrain their parameters. Because spectra for our candidate members are not available, it is not currently possible to derive a meaningful IMF for these objects.  Nevertheless, the potential to determine the IMF for this cluster from
$\sim$ 2 M$_{Jup}$ through the substellar boundary is now a step closer.


\subsection{Contamination}

\subsubsection{Extragalactic}

We have completed a deep, wide-field, near-infrared imaging survey of the $\rho$ Ophiuchi cloud core to search for candidate low mass member YSOs.  Establishing membership of a given YSO in a 
star-forming region generally requires multi-wavelength observations, since multiple indicators of youth are required to establish membership for any individual candidate source. The candidate YSOs we have identified were selected based on fits to broad-band spectral energy distribution (SEDs).
Many of our candidate objects display infrared excess emission, generally a good indicator
of youth.  Infrared excess, by itself, however, may not always be definitive to establish membership of 
an individual source, since background active galactic nuclei (AGN) could mimic YSO colors. However, given the high extinction region to which our observations were limited, the effects
of contamination by background galaxies, AGN, or red giants are minimized, as demonstrated 
below.

An upper limit to the number of extragalactic contaminants among our candidate members
can be obtained by contrasting the $K_s$ vs. $T_{eff}$ plots for sources projected within the 
cloud core in Figure 1 (the``cloud"  region), with those sources in the cloud ``exterior'' region.  Here, the ``exterior''
region is the same one as defined in \citet{MAA10}.  
Using the list of spectroscopically confirmed brown dwarfs from 
\citet{GEE11} as a guide for the location of cloud members in the de-reddened $K_s$ vs. $T_{eff}$ plot, and assuming that all of the inferred cluster members in the cloud ``exterior" region in such a plot are spurious, an estimate
of the (largest possible) number of extragalactic contaminants can be made in the following manner.

The number of objects which fall below the main sequence in the $K_s$ vs. $T_{eff}$ plots of Figure 4
and are thus identified as background stars, in the ``cloud" (including the ``excess'' and ``non-excess'' sources)
region is 404, whereas the corresponding number in the cloud ``exterior" region is 385. The number of sources above the main sequence in the ``cloud" and ``exterior" regions are 948 and 63, respectively. The number of contaminating sources in the ``cloud" region can be predicted by scaling the number of ``exterior" region objects that are above the main sequence (63) by the cloud:exterior background source count ratio, which is equal to 1.05 from the background source counts estimated above (404/385). Because background source counts are heavily affected by extinction, it is not appropriate to scale by the relative areas of the two regions. Rather, the scaling must be based on the number density ratio of extragalactic sources to background stars, which is the same for the ``cloud" and ``exterior" regions. 
Multiplying our cloud:exterior background source count ratio (1.05) by the exterior:cloud PMS source ratio (63/948), we find an upper limit to the percentage of contaminating sources among our candidate members of $\sim$ 7\%. Thus, 66 of our 948 candidate YSOs could be extragalactic background objects.

An alternative approach to estimating background contamination by extragalactic sources
may be derived by inspection of Figure 6, in which we plot the $JHK$ colors of 
galaxies in the GOODS (Great Observatories
Origins Deep Survey)-South field imaged with VLT-ISAAC, after transforming the $J-H$ and $H-K_s$ colors to the CIT system.  This makes Figure 6 directly comparable to Figure 3. 
The NIR GOODS-S data were acquired over a $\sim$ 160 arcmin$^2$ region \citep{RET10}.
The total number of galaxies is 76 in this field to our completeness limits. The GOODS-S field
was observed through negligible extinction, whereas the {\it minimum} extinction towards our 
920 arcmin$^2$ region is $A_v=5$ (see text regarding Figure 3 in $\S$3, the Results section).
Therefore, in Figure 6 we present plots of the appearance of the same 76 galaxies as if they were observed through $A_v=5$ (grey filled circles) and $A_v=10$ (black filled circles). Increasing the extinction has the dual
effect of reddening and dimming these sources, such that only 42 galaxies and 22 galaxies, respectively,
would be detected to our completion limits seen through $A_v=5$ and $A_v=10$.
More importantly, the general population of background galaxies at these faint NIR magnitudes have
very blue colors, so that they would not contaminate the ``excess'' region of the $JHK_s$ color-color plot.

\subsubsection{Galactic}

Red giants and faint red dwarfs are expected to be the major source of Galactic contaminants to our sample of candidate new, low-mass members of the $\rho$ Oph YSO population.  Such contamination is minimized given the high extinction over most of our survey area, combined with its relatively high galactic latitude (16.377799$^{\circ}\le b\le17.153386^{\circ}$).  

An excellent estimate of Galactic contamination can be made directly from our modeling efforts.
Of the 2283 total sources detected to our $JHK_s$ completeness limits, 1741 (including 830 ``excess''
and 530 ``non-excess'') sources have available {\it Spitzer} photometry.  Of these,  
827 ``excess'' and 527 ``non-excess'' sources could be successfully fit. Of the 827 successfully fit ``excess'' sources, plotted in the 
top panel of Figure 4, 61 fall below the main-sequence locus for an assumed distance of 124 pc, whereas of the 527 successfully fit ``non-excess''
sources, plotted in the middle panel of Figure 4, 343 fall below the main-sequence. Thus,
the total number of objects falling below the main-sequence from our fits is 404. This is
our estimate of the number of Galactic background stars amongst the objects for which 
good fits to our $JHK_s$ and {\it Spitzer} photometry could be made.  These Galactic contaminants
are not counted amongst our 948 candidate YSOs. 

\subsection{Comparison with Other Surveys}

A total of 316 verified (or candidate) members of the $\rho$ Ophiuchi cloud are listed in a recent review article 
\citep{WIL08}. Of these, 219 lie within our survey boundaries, and only 28 have $K_s \ge 14$.
Of the 219 objects lying within our survey
boundaries, 70 are saturated in our 
data at $K_s$ band, and a further 40 are undetected in our $J$ band data. This leaves 109 targets for which we have
full $JHK_s$ and {\it Spitzer} IRAC photometry. All 109 were fit with our model-fitting algorithm, and 
87 were found to lie above the main-sequence.  Of these 87 pre-main-sequence sources, 57 lie
in the ``excess'' region of the $J-H$ vs. $H-K_s$ diagram, and 30 are in the ``non-excess'' region.
This result demonstrates the efficacy of our modelling
method at identifying pre-main-sequence objects in an unbiased fashion, with regard to the presence or
absence of disks. 

{\it Spitzer}-selected objects, without available corresponding deep NIR photometry,
are necessarily biased towards identifying sources with disks, often necessitating follow-up with
X-ray telescopes to identify the ``missing'' disk-less populations \citep[e.g.,][]
{BAR11, PIL10, WIN10}.  Assuming the availability
of sufficiently deep NIR and mid-IR photometry, our method presents an alternative approach to identifying
the disk-less population in nearby star-forming regions.  {\it This method also provides, for the first time,
a uniform, unbiased means for identifying the entire pre-main-sequence population in these regions, in 
a statistical sense}. 

Of course, selection biases are inherent in any observational effort--the aim is to 
understand what inherent biases there may be and to minimize their effect. Clearly, our method
would not detect the two known Class 0 objects (VLA1623 and IRAS 16253$-$2429) in L1688. 
Very faint, nearly edge-on disk systems might also be missed, due to lack of detection at $J$-band.
Nevertheless, the efficacy of this method for detecting a heretofore undiscovered, significant new pre-main-sequence
population has been demonstrated, and the application of this method for uncovering new populations
in other nearby star-forming regions is ongoing.  Quantitative evaluation of selection biases inherent
in this method, vis-a-vis evolutionary stage, source orientation, and degree of embeddedness in the cloud
awaits future work.

A comparison of our $JHK$ photometry with published photometry from the three recent, deep NIR surveys of $\rho$ Oph
is presented in Table 4. The coordinates of each source are listed first
(as determined from our astrometry, described in $\S$2 above),
followed by our $JHK_s$ photometry.  The next set of columns display Alves
de Oliveira et al's (2010) source identifications and 
$JHK_s$ photometry for sources in common with our survey (from their Table 4).
The next set of columns display Marsh et al's (2010a)
source identifications and $JHK_s$ photometry for sources in common with
our survey (from their Table 1).  Finally, the last 3 columns of Table 4 display Geers et al's (2011)
source identifications and $JK_s$ photometry for sources in common with our survey
(from their Tables 1 \& 2).  Graphical presentations of these photometric
comparisons are displayed in Figure 7.

\citet{ALV10} list 110 candidate sub-stellar objects,
of which 74 fell within our survey area.  We list photometry for all but two
of these (their Source 16, which fell on bad pixels, and 
their Source 72, which lies on a bright diffraction spike in our images).
The photometric agreement between the two datasets
is generally good (see top left column of Figure 7), with only Source 30
exhibiting highly discrepant values. 


There are only 8 sources in common between Alves de Oliveira's Table 4 and Marsh et al's Table 1.  
These correspond with Marsh et al's source nos. 829, 311, 654, 2978, 313, 222, 239,  and 334.
The root mean square error between the two sets of photometry for these sources, excluding Source 2978, is within 0.18 magnitudes at $J$, 0.12 magnitudes at $H$, and 0.090 magnitudes at $K_s$.
For Marsh et al's Source 2978, the magnitude differences between the two sets of photometry
are 2.66 at $J$, 0.90 at $H$, and 0.21 at $K_s$, with Alves de Oliveira's values always fainter.
For the same source, our photometry varies from Marsh et al's by 0.29 at $H$ and 0.08 at $K_s$, with our values being the fainter ones (this source was not detected in our data at $J$). This photometric discrepancy
may be due to the presence of extended faint nebulosity surrounding this object, as elaborated in
the following for the case of the spectroscopically confirmed planetary mass object, Marsh et al's Object 4450 \citep{MAB10}.

For the seven candidate planetary mass objects for which spectra were obtained by \citet{MAB10}, Alves de Oliveira state they found good agreement (between 0.02 and 0.23 magnitudes difference)
between the two sets of photometry at $K_s$ for Marsh et al's source nos. 
1449, 1307, 2438, and 2403, but differences 
of 0.4, 1.42, and 1.47 magnitudes at $K_s$-band for Marsh et al's source 
nos. 2974, 4450, and 3117, respectively.  Unfortunately, Alves 
de Oliveira et al. did not publish the actual values for their $JHK_s$ magnitudes for any of these sources,
except for the $J$ and $K_s$ values for the spectroscopically confirmed planetary mass object, Object 4450.
Our $K_s$ photometry agrees with that of Marsh et al. within 0.1 mag for Source 1307, 
within 0.21 mag for Source 2438, 
and within 0.02 mag for
source 2403, 
in agreement with Alves de Oliveira's stated range of magnitude differences for these sources.
Our $K_s$ value for Object 4450 is 18.15, 0.44 fainter than Marsh et al's $K_s$ value of 17.71, but 1.01 magnitudes
brighter than Alves de Oliveira's $K_s$ value of 19.14. Alves de Oliveira et al. could not derive an $H$-band value
for Object 4450, due to an image artifact in their data. We derive $H$=18.76 compared with Marsh et al's value of $H$=18.36.
We did not detect Object 4450 at $J$ band, whereas Marsh et al. derived $J$= 19.57 and Alves de Oliveira derived
$J$=21.32 $\pm$ 0.35. Based on the large discrepancy between their $K_s$ band value from Marsh et al's for Object 4450,
Alves de Oliveira et al. suggest that this planetary mass object lies as far as the Sco-Cen association, and is not associated with
the $\rho$ Oph cloud.  Our results do not support this conclusion.

There is a possible scenario that would resolve the issue of such large photometric discrepancies being
reported by different workers for the planetary mass Object 4450.
Note that the observed large differences in photometry between Marsh et al. and Alves de Oliveira et al.
occur for 4/15 sources, suggesting the presence of a systematic, rather than random measurement
error. One possibility is that of extended emission surrounding the objects with discrepant photometry, which we argue to be the case for Object 4450.
The pixel size of 2MASS was 2$^{\prime\prime}$ (used for the Marsh et al. photometry);  the pixel size reported 
in this work was 0.45$^{\prime\prime}$ (albeit in 1.0-1.4$^{\prime\prime}$ seeing), and the pixel size for the CFHT WIRCam
observations was 0.3$^{\prime\prime}$, in excellent seeing, ``typically between 0.4$^{\prime\prime}$-0.5$^{\prime\prime}$'', but always
better than 0.8$^{\prime\prime}$. Alves de Oliveira used PSF-fitting photometry, which would miss measuring any extended flux, and
would result in fainter measured magnitudes than would be derived for an extended object from aperture photometry. 
The hypothesis of an extended source is supported by our measurements falling between the values found by 
Alves de Oliveira on the one hand, and Marsh et al. on the other.

Of the 165 sources listed in Table 1 of \citet{MAA10}, photometry for 92 are listed in Table 4.
Of the 73 sources not listed in Table 4, 50 are outside of our survey area, 20 are saturated in our data and 3 are below our detection threshold.
The general photometric agreement between the two datasets is satisfactory and
is plotted in the middle column of Figure 7.

\citet{GEE11} list 36\footnote[10]{Entries 8 and 24 in Table 1 of
\citet{GEE11} are identical.}
 ``likely substellar members with disks'' in $\rho$ Oph in their Table 1, of which 10 have NIR photometry from MOIRCS and 27 have NIR photometry from 2MASS.
They list a further three ``Probable Low-Mass and Substellar Members of $\rho$
Oph with MOIRCS Spectroscopy Follow-up'' in their Table 2.  We recover all but four sources of which three (their Table 1 entries
11, 15, and 27) were outside of our survey limits, and one (their their Table 1, entry 10)
which was too faint at J to be detected by our survey, and is located in regions of very bad signal to noise in our survey at $H$ and $K$.  The photometric agreement between our surveys at $J$ and $K_s$
(no $H$ band data were acquired by \citet{GEE11}) is quite good as can be seen from the plots in the right column of Figure 7.
 
Recently, \citet{GEE11} have estimated an upper limit for the ratio of low-mass stars 
(0.1 M$_{\odot} \le$M$\le$1.0 M$_{\odot}$) to brown dwarfs (M $\le 0.1$M$_{\odot}$) in the $\rho$ Ophiuchi cloud to be $\sim$ 3 - 7. 
An alternative upper limit to the low-mass star: brown dwarf ratio in $\rho$ Oph 
can be derived using
the subsample of ``non-excess'' sources lying above the main-sequence (as plotted in the
middle panel of Figure 4). 

Only the 
``non-excess'' sources whose SED fits had fluxes
within a factor of 3 of what a photospheric model would predict at the distance of $\rho$ Oph
are considered, in order to exclude YSOs with non-photospheric mid-IR emission that might
masquerade as having low-temperature photospheres (see discussion of this point in $\S$4.1).
Mass estimates for these sources are derived from their best-fit, 1MYr COND and DUSTY models, 
yielding 59
objects with masses in the range 0.1$\le$M$\le$1.0M$_{\odot}$, and 
83 objects
with M $\le 0.1$M$_{\odot}$, thus yielding a value of $\sim$ 0.7 for the low-mass star: brown dwarf
ratio.  However, unaccounted for systematic biases ({\it e.g.,} under-estimating the number
of higher-mass PMS sources) may have entered into this estimate. The point is to illustrate the
variation in the possible range of this value in $\rho$ Oph, given the present data.
\citet{MAA10} have recently published an estimate of the IMF in the $\rho$ Ophiuchi cloud. Their results show an increase in the number of cloud members progressing from 0.1 M$_{\odot}$ to lower masses.  Our results, so far, are consistent with theirs.  However, it would be premature to derive a definitive IMF across the substellar boundary 
from the data presented here.
Future spectroscopy of our candidate cloud members will allow the construction of an IMF for this region, and further refine the value for the low-mass star: brown dwarf ratio in $\rho$ Ophiuchi.



\section{Summary and Conclusions}
\begin{itemize}

\item We present a new, deep, $JHK_s$ survey of a
920 arcmin$^2$ area of the $\rho$ Ophiuchi star-forming cloud,
encompassing its highest extinction core to 90\% completeness
limits of $J=20.0$, $H=20.0$, and $K_s=18.50.$  Our survey
is thus sensitive to an object of just $\sim$ 1.5 $M_{Jup}$ with
an age of 1 MYr and photospheric temperature of $\sim$1100 K
at the distance to $\rho$ Oph, This mass sensitivity falls to
2.0, 4.0, and 8.5 $M_{Jup}$ for $A_V=5$, 10, and 15, respectively.

\item We combine our new, deep, $JHK_s$ photometry with mid-infrared
{\it Spitzer} photometry to produce SEDs for a total of 1741 sources
within our survey boundaries. These sources are divided into three groupings according
to their placement in the $J-H$ vs. $H-K_s$ color-color diagram into:
{\it i)} 830 ``excess'' sources, those which de-redden to the CTTS locus of 
\citet{MEY97};
{\it ii)} 533 ``non-excess'' sources, which fall within the main-sequence reddening band, 
and {\it iii)} 378 sources that can be de-reddened to the colors of M7$-$L0
spectral types with a resulting large variation in their deduced extinction values.

\item An improved version of the fitting procedure of \citet{MAA10} is used
to fit atmospheric models (COND, DUSTY, NextGen) and single-temperature blackbody
spectra to the observed SEDs for an age of 1 MYr
and a distance of 124 pc, appropriate for the age and distance of the $\rho$ Oph cloud's YSO population.  

\item Of the 827 successful SED fits for ``excess'' sources, 764 are found to lie above the main-sequence.
Of the 527 successful SED fits for ``non-excess'' sources, 184 lie above the main-sequence.
We therefore identify 948 candidate pre-main-sequence sources, of which 87 (57 ``excess'' and 30
``non-excess'') are duplicates with sources listed in \citet{WIL08}.

\item The fact that 184 ``non-excess'' and 764 ``excess'' sources are identified as pre-main-sequence
demonstrates the efficacy of this method for identifying the entire pre-main-sequence population in the surveyed area, unbiased with respect to the presence or absence of disks.

\item Of the 378 sources with complete $JHK_s$ and {\it Spitzer} SEDs that can be
de-reddened to the colors of M7$-$L0 photospheres in the $J-H$ vs. $H-K_s$ color-color
diagram, the fraction above the main-sequence
varies from 22\% (82 above main-sequence/374 successful SED fits) if all such sources are de-reddened to M7 colors, to 95\% (357/377) if all such sources are de-reddened to L0 colors, and
78\% (294/377) for sources de-reddened to colors that are the average between M7 and L0 colors
(close to M9, in practice).  Follow-up spectroscopy is required to decide what fraction of this sample
represents further augmentation of the pre-main-sequence population of the cloud, and what 
fraction are reddened background stars. 

\item The embedded population of candidate YSOs in the $\rho$ Oph core is increased
by a factor of $\sim$ 4 by this study, 
even allowing for contamination of the cluster member candidate sample by
background galaxies or AGN.

\item Follow-up spectroscopy of the cluster member candidates opens up the possibility
for determination of the IMF in this star-forming cloud throughout the brown dwarf
mass range, reaching well into the planetary mass regime. 

\end{itemize}

\acknowledgements

We thank the AAO staff for their outstanding support in making our observations possible. We also thank Utah Valley University undergraduate physics student, Sherene Higley, for her assistance in the reduction of the AAO data. M.B., K.H., and C.M. acknowledge the support of 
NSF Research at Undergraduate Institutions grants AST-1007928, AST-1009776, and AST-1009590, respectively, for support of this research.
M. B. gratefully acknowledges NSF grant AST-0206146 which made further contributions to this work possible. Additional support for this work was provided by the National Aeronautics and Space Administration through Chandra Award Number AR1-2005A and AR1-2005B issued by the Chandra X-Ray Observatory Center, which is operated by the Smithsonian Astrophysical Observatory for and on behalf of NASA under Contract NAS8-39073.

\clearpage

\begin{figure}
\centering
\includegraphics[height=7in]{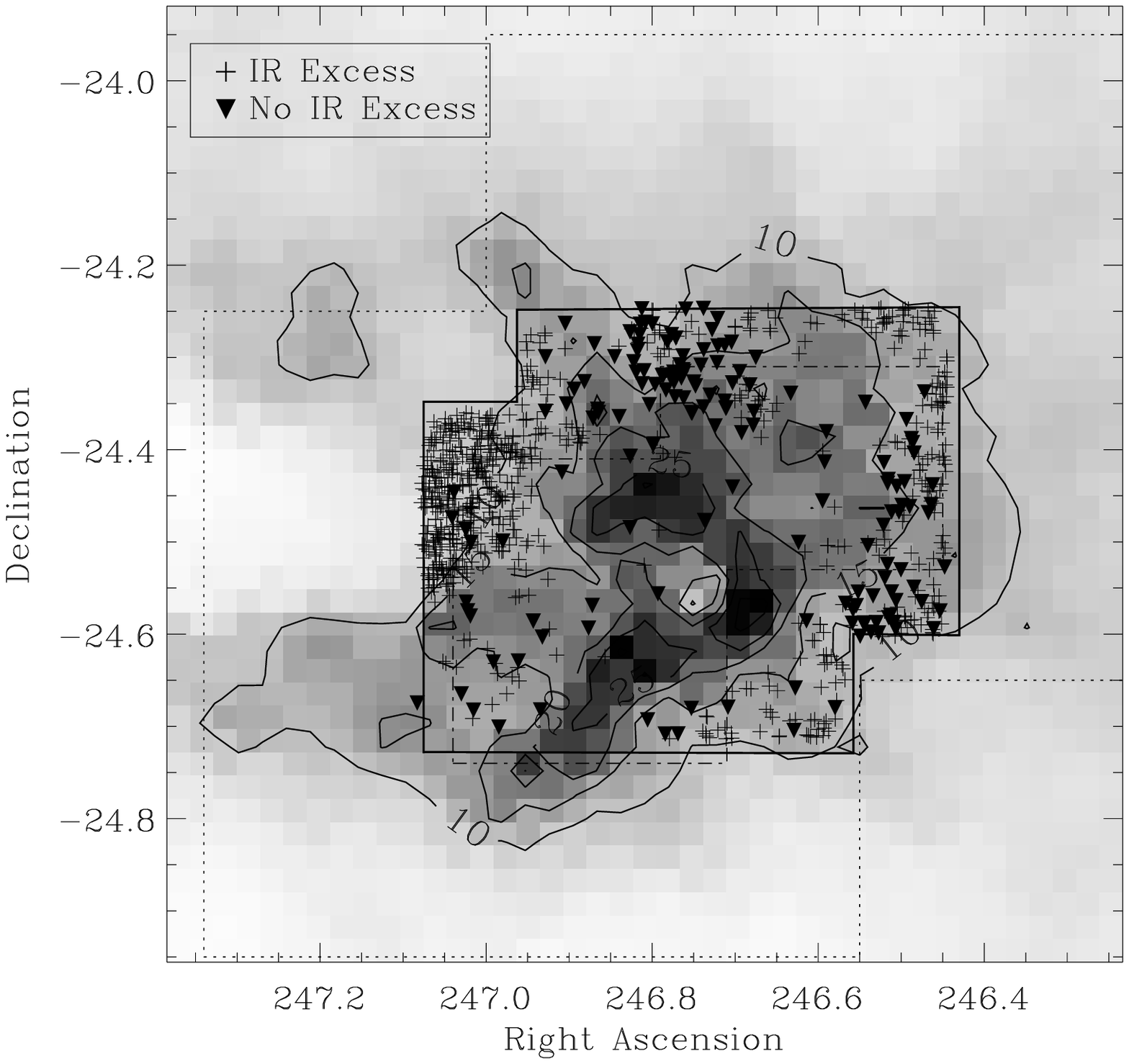}
\caption{Plot of spatial distribution of our candidate $\rho$ Oph members (crosses$=$excess sources, filled triangles$=$non-excess sources) superposed
on the $A_V$ contours from the COMPLETE project.  $A_V$ contours are plotted and labelled at $A_V=10, 15, 20,$ and 25. Our 920 arcmin$^2$ survey
area is indicated by the solid outline. The survey areas of Geers et al.
and Alves de Oliveira et al. are indicated by the dot-dashed and dotted outlines, respectively.\label{figure1}}
\end{figure}

\begin{figure}
\centering
\includegraphics[height=2in]{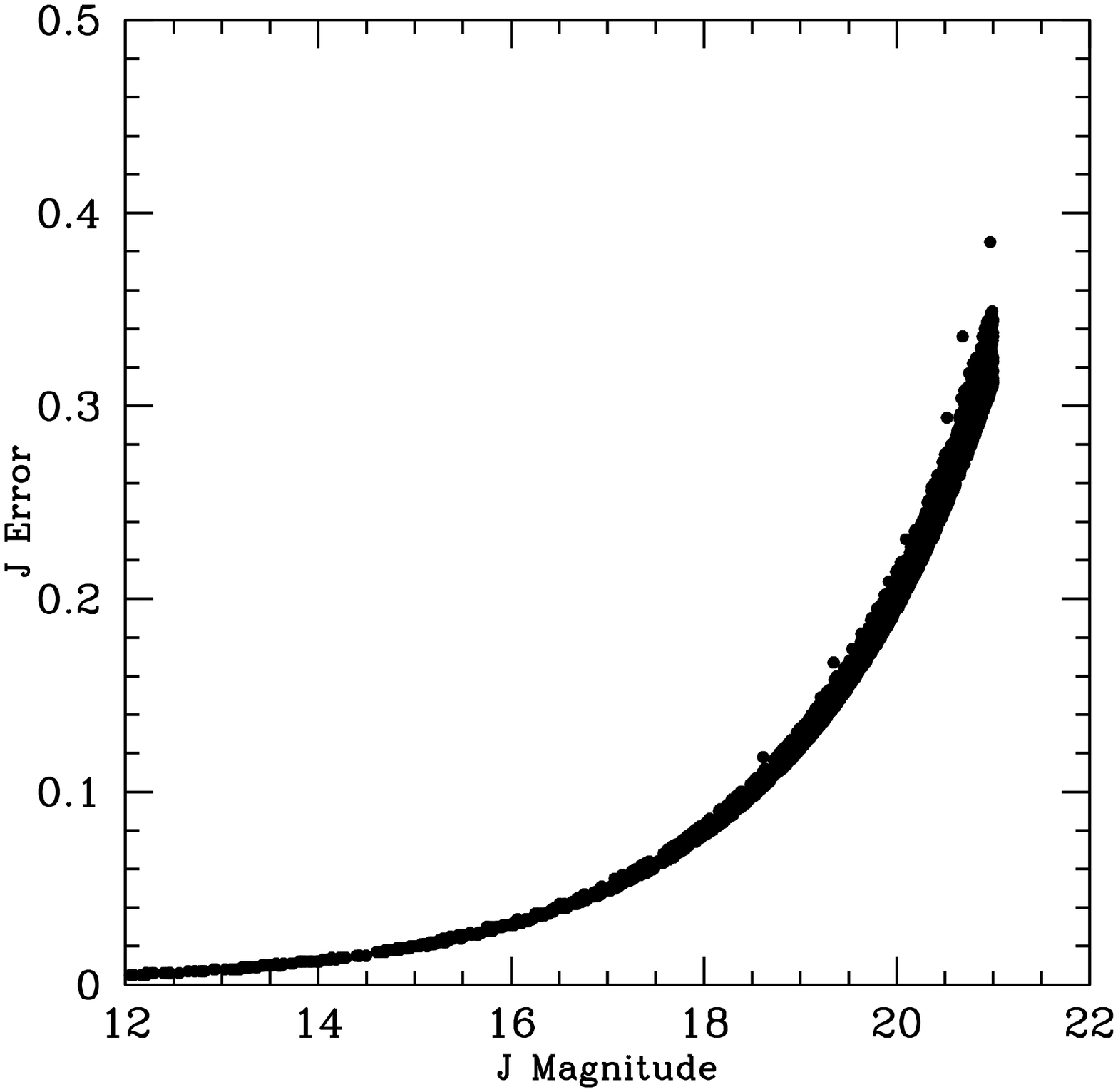}
\includegraphics[height=2in]{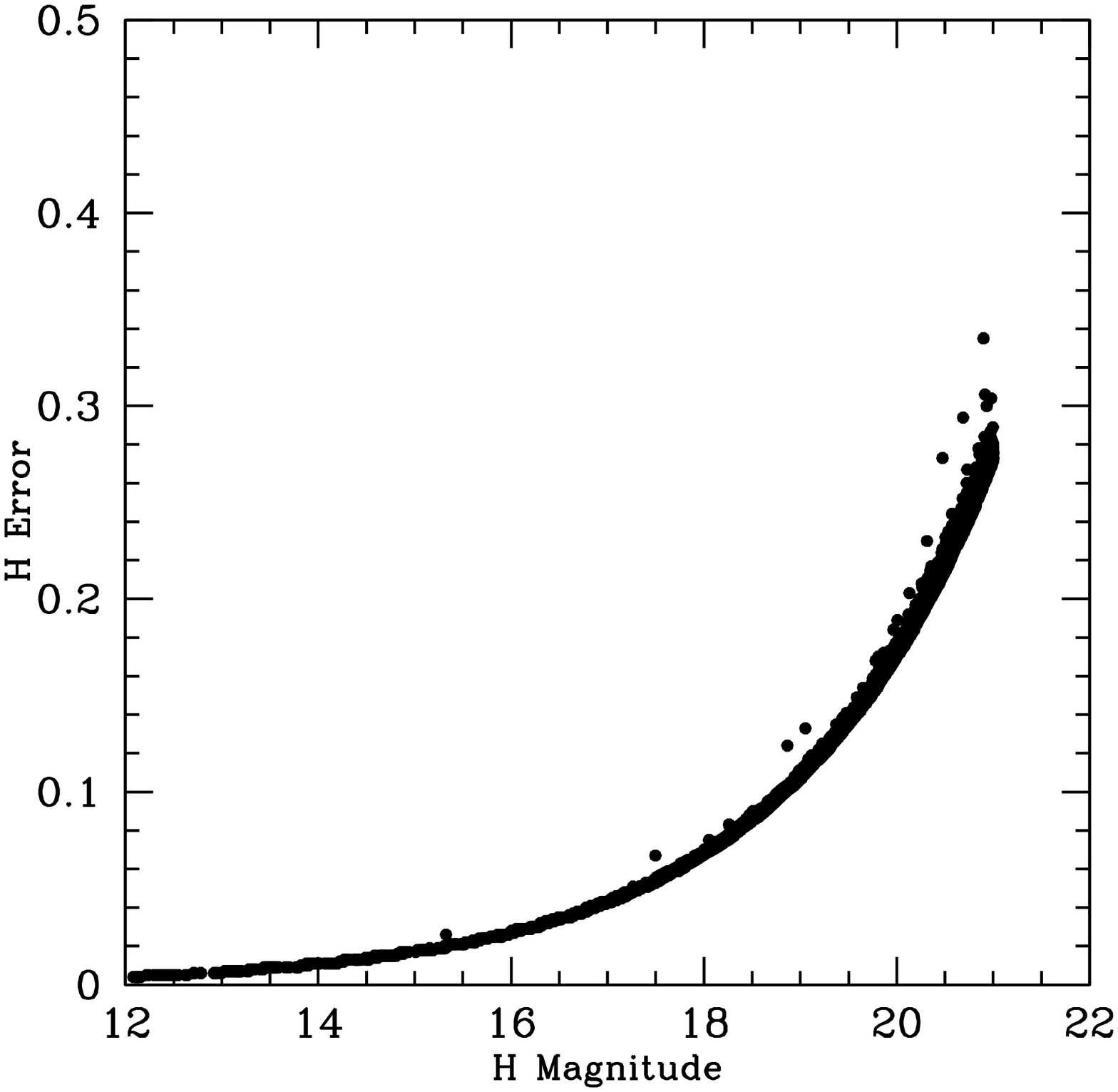}
\includegraphics[height=2in]{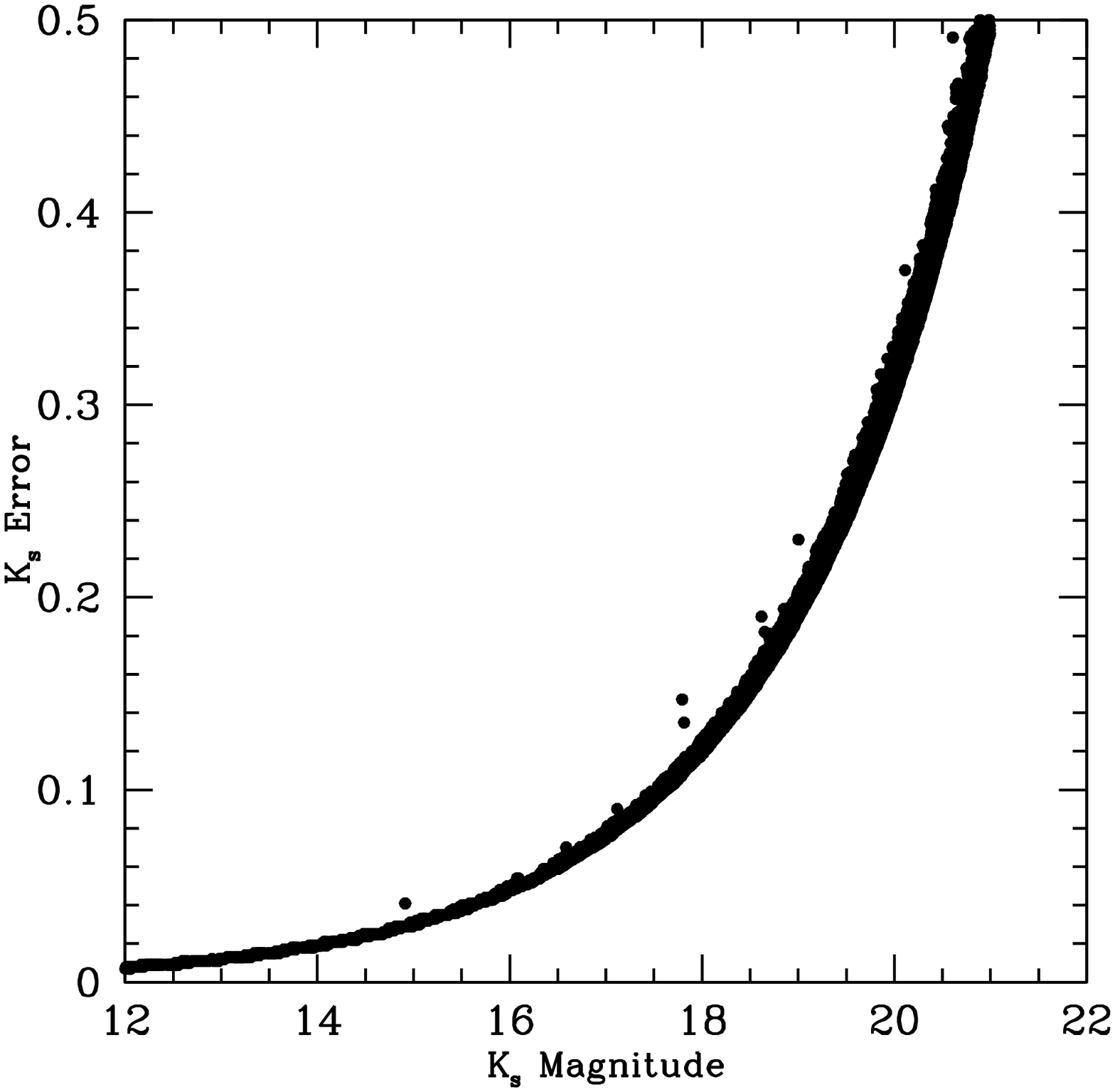}
\caption{Photometric errors as a function of magnitude at $J$ (left panel),
$H$ (middle panel), and $K_s$ (right panel). Completeness limits
are at $J=20.0$, $H=20.0$, and $K_s=18.50$, as discussed in the text.
Saturation limits are at $J\simeq 12.0$, $H\simeq$11.0, and $K_s\simeq10.0$.
\label{figure2}}
\end{figure}

\begin{figure}
\plottwo{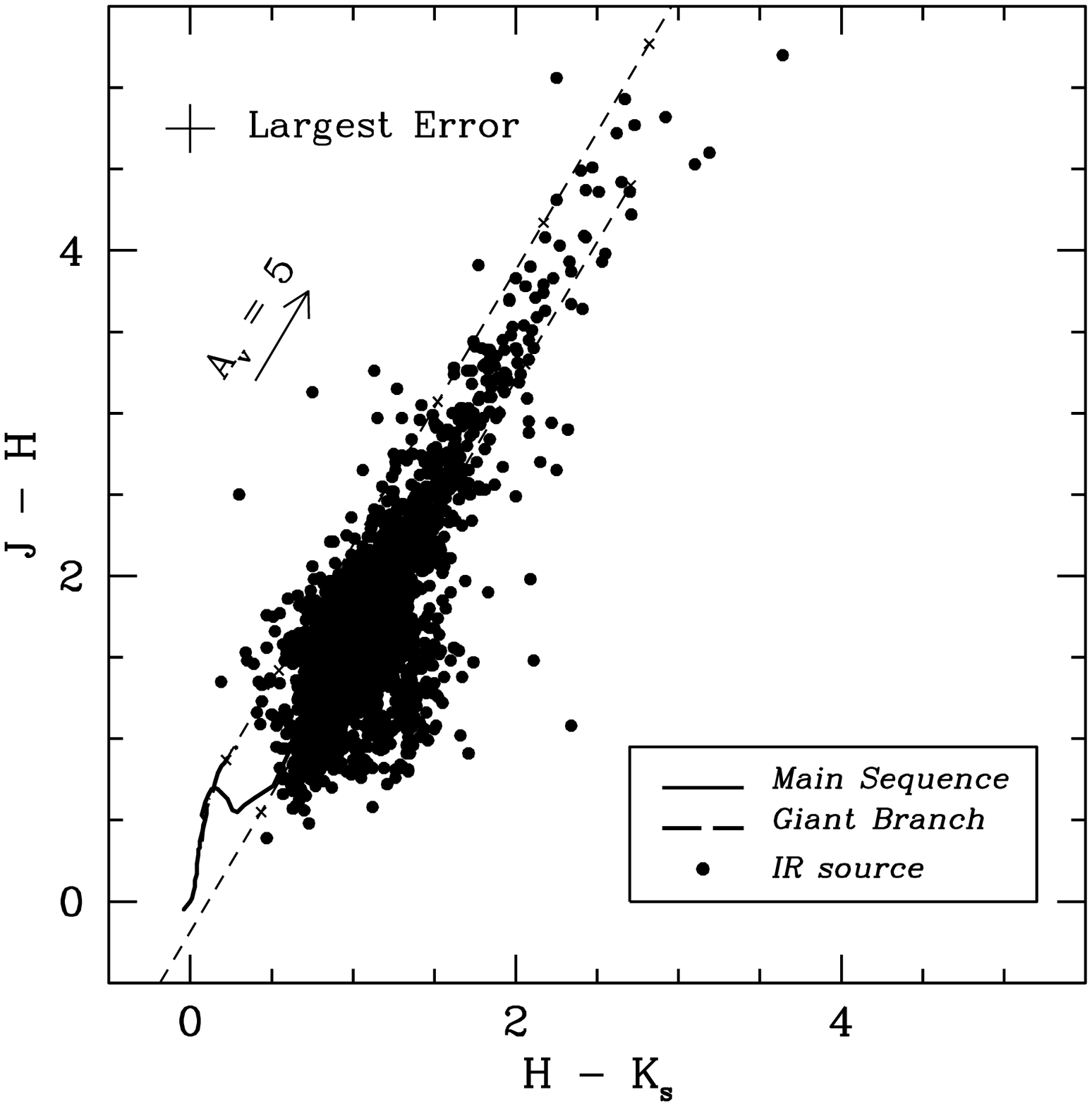}{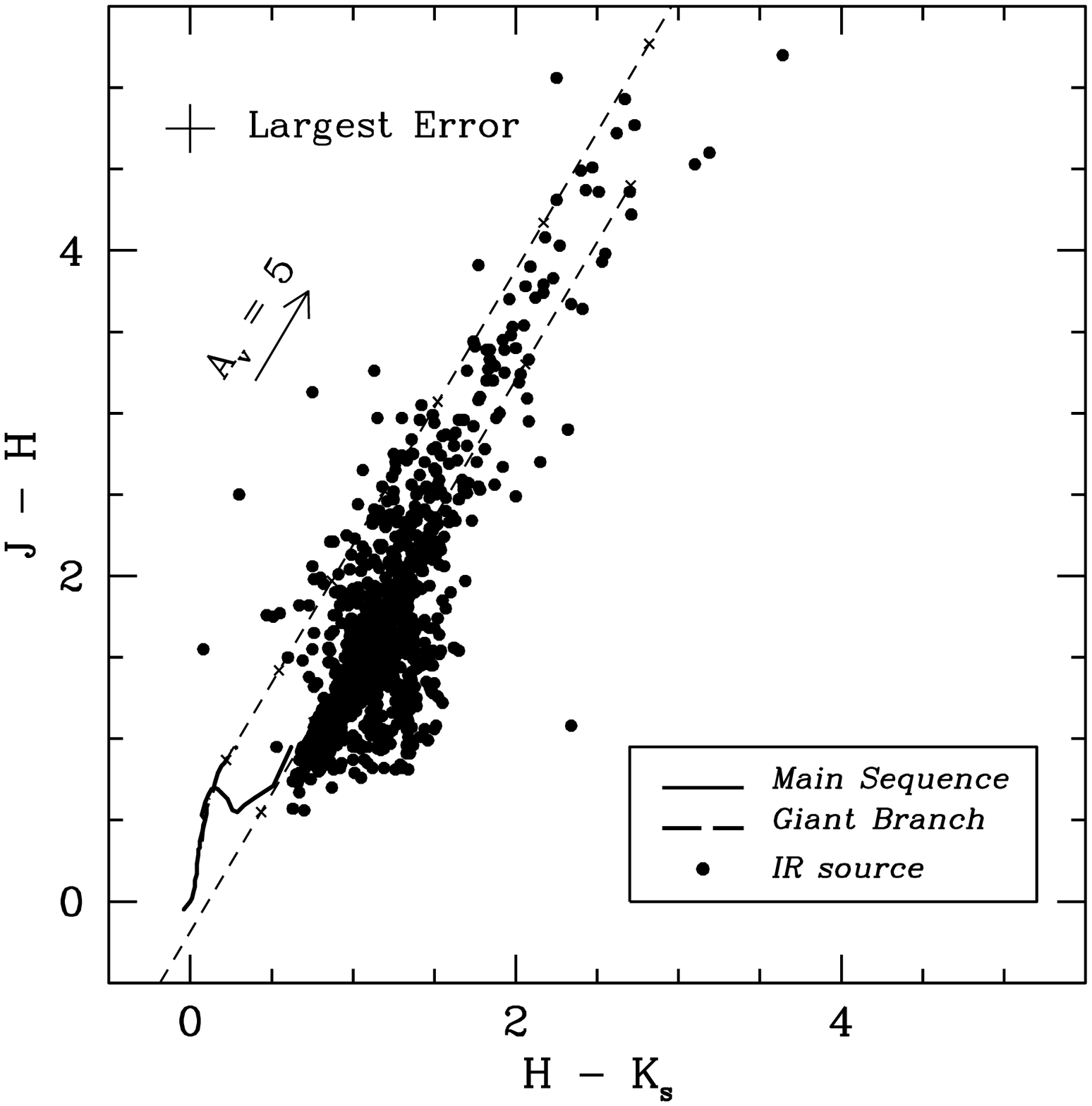}
\caption{$JHK_{s}$ color-color diagrams for sources with 10 $\le K_s \le$ 18.5 detected in all 3 bands
in our survey of $\rho$ Oph {\em (left)}, and for all objects falling above the main-sequence after SED-fitting (see text) {\em (right)}. In the diagrams, the solid line represents the locus of points corresponding to the unreddened main sequence, continuing into the realm of cool young photospheres.  The dashed diagonal lines indicate the main-sequence reddening band.
The locus of positions of giant stars is shown as a heavy dashed line. The CTTS locus (not plotted)
extends from [0.81,0.50] $\le$ [$J - H$,   $H - K$] $\le$ [1.10, 1.00]. The diagonal arrow represents the effect of 5 magnitudes of visual extinction. The uncertainty in the colors for all sources is magnitude-dependent,
but always $\le$ 0.2 magnitudes, as labelled in both diagrams.\label{figure3}}
\end{figure}

\clearpage

\begin{figure}
\centering
\includegraphics[height=6in]{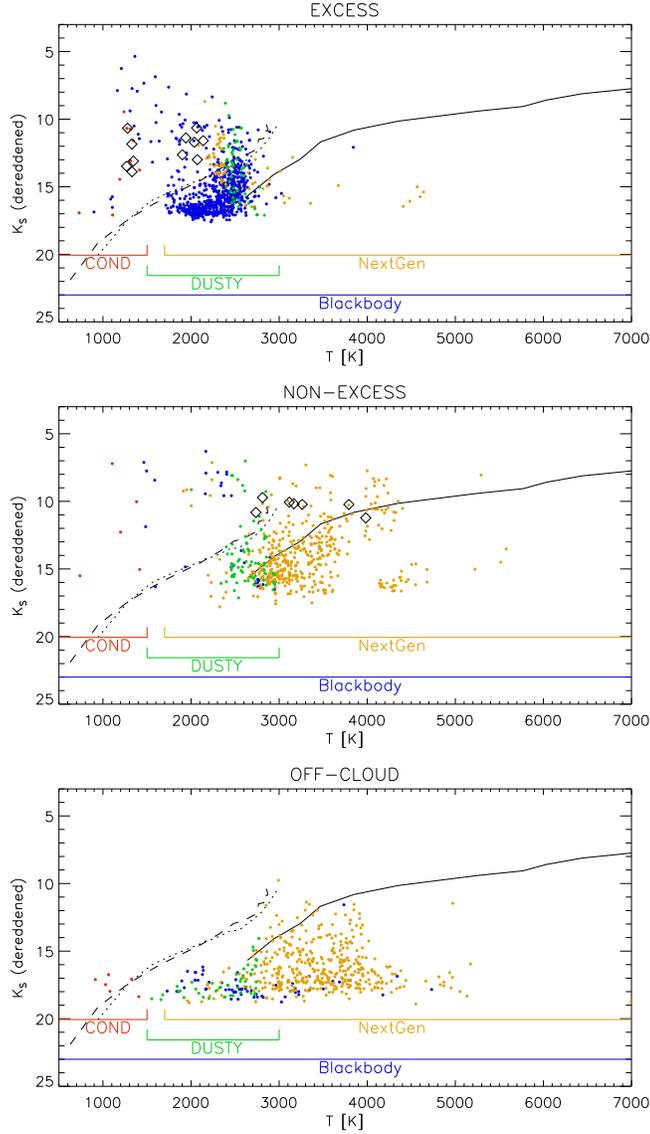}
\caption{
Dereddened $K_{s}$ magnitude as a function of estimated temperature for 
sources for which successful SED fits were obtained to their $JHK_{s}$ and {\it Spitzer} IRAC 
data.
In the top panel, we plot the locations of the 827 ``excess'' sources, along with the locations of
known, spectroscopically confirmed brown dwarfs with IR excesses, which are indicated by open
diamond symbols.
In the middle panel, we plot the locations of 527 ``non-excess'' sources ,
along with the locations of spectroscopically confirmed brown dwarfs lacking IR excesses,
which are indicated by the open diamond symbols.
In the bottom panel, we plot the locations of sources located in the cloud-exterior region from \citet{MAA10}. Model curves are plotted in each panel for the 1 Myr COND (dashed) and DUSTY (dotted) models, as well as for the  main-sequence NEXTGEN models (solid), for an assumed distance of 124 pc.\label{figure4}}
\end{figure}

\clearpage


\begin{figure}
\centering
\includegraphics[height=5in]{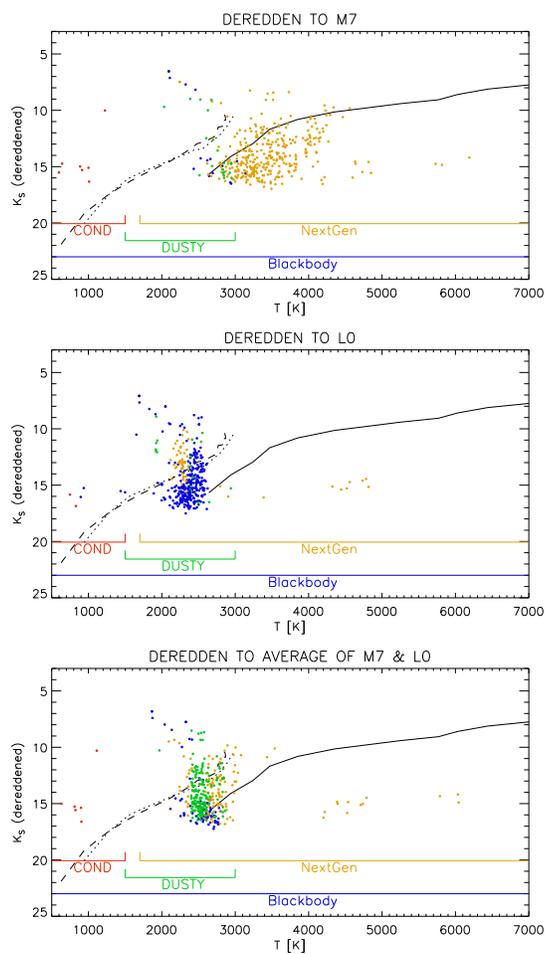}
\caption{Same as Fig. 4, except for ``non-excess'' sources which de-redden to low-mass photospheric colors.  The effect of assuming intrinsic $J-H$ vs. $H-K_s$ colors of M7 (top panel),
L0 (middle panel), or to an average of M7-L0 spectral types (bottom panel) results in 
extinction values different enough to cause the demonstrated variation in the distribution
of the best-fit SEDs in de-reddened $K_s$ vs. $T_{eff}$ space.  \label{figure5}}
\end{figure}
\clearpage

\begin{figure}
\centering
\includegraphics[height=7in,angle=90]{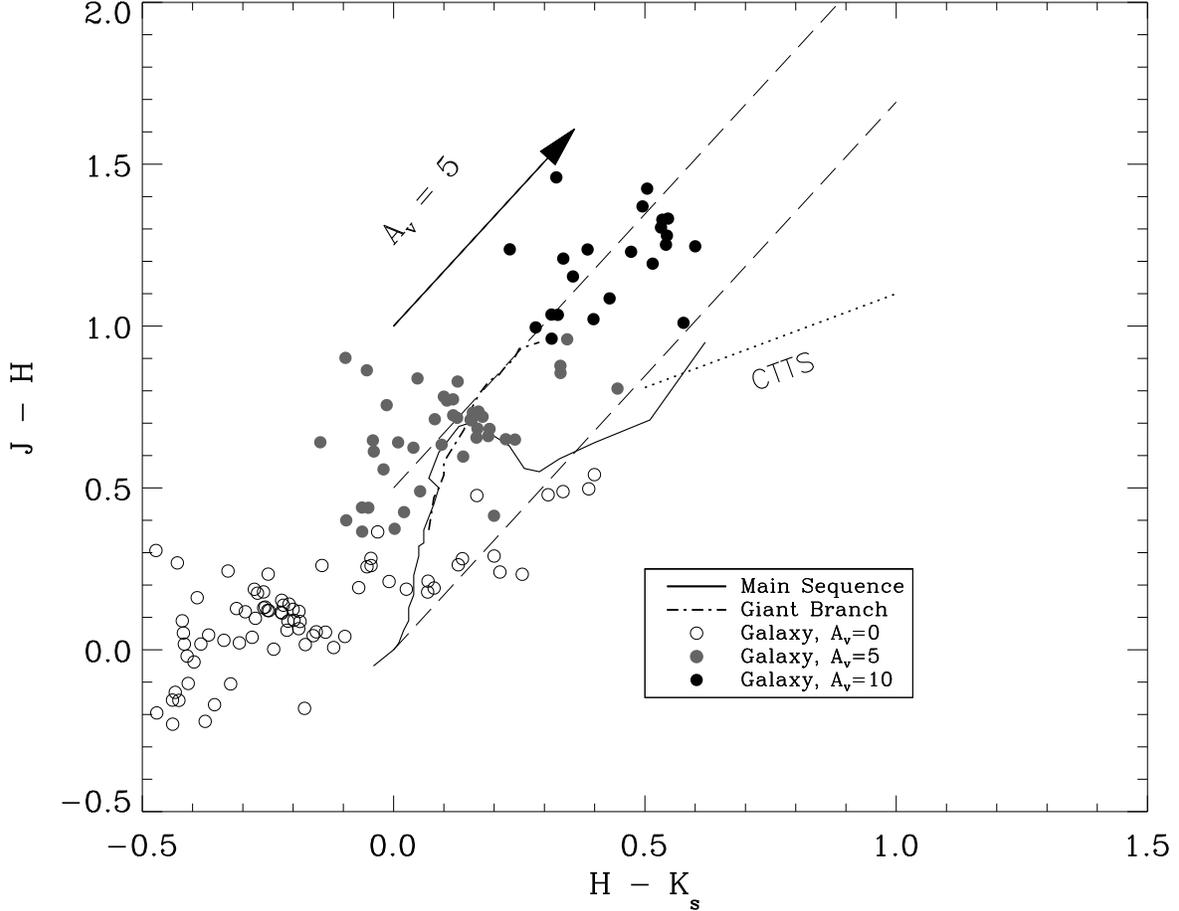}
\caption{
$JHK_{s}$ color-color diagrams for all galaxies from the GOODS-South field to our completeness
limits of $J=20.0,\ H=20.0,\ K_s=18.50$ with no extinction (open circles), seen through $A_V=5$ (grey circles), and seen through $A_V=10$ (black circles). The loci of old main-sequence stars and substellar objects (solid curve), giants (dot-dashed line) , and the Classical T-Tauri Star (CTTS) locus (dotted line)
are all shown for clarity. The parallel dashed lines delineate the reddening band--objects in this 
region de-redden to main-sequence or giant colors.
Comparison with Figure 3 shows that only a small fraction of background galaxies would fall into the ``excess'' region to the right of the right-most reddening line.\label{figure6} }
\end{figure}
\clearpage

\begin{figure}
\plotone{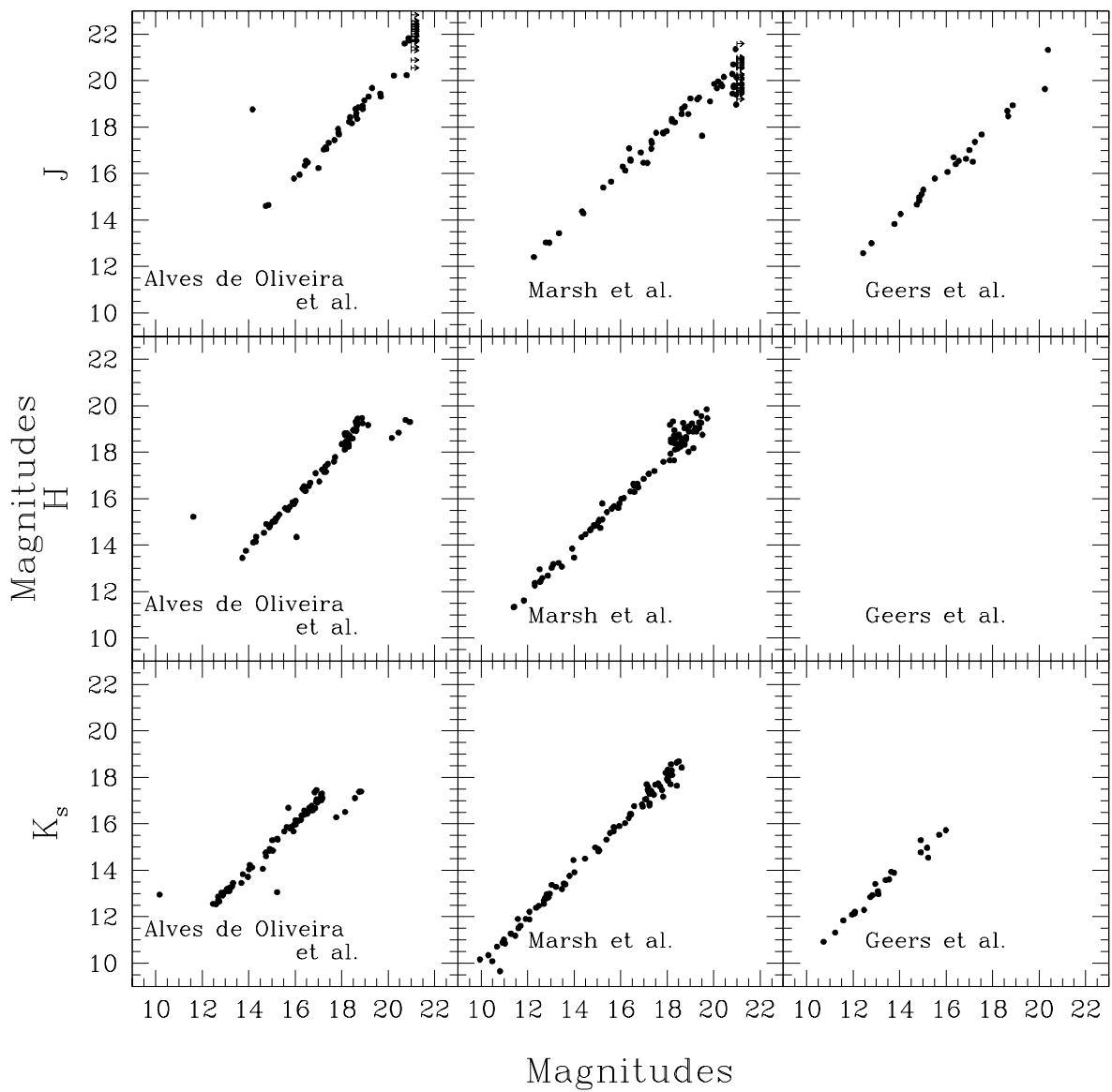}
\caption{Comparison between photometry presented in this work (horizontal axes)
with photometry published by other workers (vertical axes) for sources in common as presented in Table 4. Our non-detections at $J$-band
are indicated by the rightward pointing arrows.  No data are plotted at $H$-band for \citet{GEE11}, since they did not obtain $H$-band photometry. \label{figure7}}
\end{figure}

\clearpage
\begin{deluxetable}{lcccccccc}
\footnotesize
\tablecaption{Nearby SFRs with Published IMFs Using NIR Imaging Data\label{table1}}
\tablewidth{0pt}
\tabletypesize{\scriptsize}
\tablehead{ \colhead{SFR} & \colhead{Distance} & \colhead{Age} & \colhead{Telescope} &
  \colhead{FOV} & \colhead{$J$} & \colhead{$H$} & \colhead{$K_s$} & \colhead{Ref.}\\
                                           &  \colhead{(pc)}       &\colhead{(MYr)} &    IR Camera            &
                        & \colhead{(mag)}& \colhead{(mag)}&  \colhead{(mag)}&            \\} 
\startdata
              &        &    &                                      &           &    &    &            &      \\
IC 348 & 300  & 2    &KPNO 2.1m                  & 20.5$^{\prime}\times20.5^{\prime}$&18.82   & 18.04 & 17.72& (1) \\
       &      &      &FLAMINGOS                  &                                   &        &       &      &     \\
       &      &      &Calar-Alto 3.5m            & 18.6$^{\prime}\times18.4^{\prime}$&19.5    & \ldots\tablenotemark{a}&\ldots\tablenotemark{a}& (2) \\
       &      &      & Omega Prime               &                                  &         &       &      &      \\  
ONC    & 430  & 0.8$\pm$0.6  &NTT 3.5m/SOFI $+$          &  5$^{\prime}\times5^{\prime}$    & 18.15   & 18.7  & 17.5 & (3) \\
       &      &      &FLWO 1.2m/STELIRcam        &                                  &         &       &      &      \\
Cha I  &160-170   & $\sim$2  &2MASS   $+$             &  1.5$^{\circ}\times0.35^{\circ}$  &15.75         &   15.25   & 14.3\tablenotemark{b}     &  (4)   \\ 
         &                &               &CTIO-4m/ISPI              &  13.2$^{\prime}\times16.8^{\prime}$  &   18.5      & 18.25      &    \ldots \tablenotemark{c}   &   \\      
         &                &               &                                 &                                                      &          &       &      &     \\
Taurus&140  & 1-1.5\tablenotemark{d}   &  2MASS    & 2.84\,${\rm deg}^2$& 15.75  & 15.25 &14.3\tablenotemark{a}  & (5)     \\                                                        
             &       &      3\tablenotemark{e}   &   2MASS  & 1.32\,${\rm deg}^2$&             &           &                                      &          \\
$\sigma$ Ori & 360$^{+70}_{-60}$ & 3$\pm$2 & VLT 8.2-m  & 790 arcmin$^2$& 20.6   & \ldots\tablenotemark{a}  &\ldots\tablenotemark{a}& (6) \\
                     &                                &                &    ISAAC      &                          &           &             &        &     \\     
                     &                                &                & UKIRT 3.8-m& 0.78 deg$^2$  & 19.0   & 18.4    & 18.0& (7) \\
                     &                                &                &    WFCAM      &                          &           &             &        &     \\     
                   
$\rho$ Oph   & 124                         & 1             &   2MASS     &1$^{\circ}\times9.3^{\prime}$& 20.5&20.0&19.0& (8) \\  
                     &                                &                &                    &                                             &        &       &       &     \\   
$\rho$ Oph   & 124                         & 1             &   AAO-3.8m &920 arcmin$^2$   & 20.0&20.0&18.5& (9) \\  
                     &                                &                &    IRIS2        &                 &         &        &        &     \\                          
                       
\enddata
\tablenotetext{a}{No data acquired at this wavelength}
\tablenotetext{b}{2MASS $K_s$ completeness from\\
 {\tt http://www.ipac.caltech.edu/2mass/releases/allsky/doc/sec2\_2b.html} }
\tablenotetext{c}{Data acquired, but $K_s$ completeness limit not published}
\tablenotetext{d}{For field centered at J2000 $\alpha$=4$^h$39$^{m}$00$^s$, $\delta$=25$^{\circ}$46$^{\prime}$00$^{\prime\prime}$}
\tablenotetext{e}{For field centered at J2000 $\alpha$=4$^h$55$^{m}$00$^s$, $\delta$=30$^{\circ}$24$^{\prime}$30$^{\prime\prime}$}
\tablerefs{
(1) \citet{MUE03}; 
(2) \citet{PRE03};
(3) \citet{MUE02}; \\
(4) \citet{LUH07};
(5) \citet{LUH04};
(6) \citet{CAB07};
(7) \citet{LOD09}; \\
(8) \citet{MAA10};
(9) This work}
\end{deluxetable}

\clearpage

\begin{deluxetable}{lcccc}
\footnotesize
\tablecaption{Substellar Objects in Nearby SFRs\label{table2}}
\tablewidth{0pt}
\tabletypesize{\scriptsize}
\tablehead{ \colhead{SFR} & \colhead{B.D./Star} & \colhead{Ref.} & \colhead{No. of}       &  \colhead{Ref.}\\
                                           & \colhead{Ratio}       &                         & \colhead{Candidate}&                         \\
                                           &                                &                         & \colhead{PMO's\tablenotemark{a}}     &                         \\                    
                                           &                                &                         &                                  &     } 
\startdata
                                           &                                &                         &                        &                                    \\
IC 348                                 &    15\%$-$25\%      &      (1,2)               &             1      &      (3)                        \\
                                           &                                &                         &                        &                \\
ONC                                   &  30\%                      &  (4,5)                &    10               & (6)          \\   
                                           &  54\%                      & (7)                    & 142-421        & (7)         \\   
                                           &                                &                         &                          &          \\                                                        
NGC1333                           &   67\%                     &     (8)                &             0          &     (8)      \\
                                           &                                &                         &                          &           \\
Cha I                                  &  17\%                      &    (9)                &              10          &  (9)          \\ 
                                           &                                &                         &                          &          \\
Taurus                               & 17\%                       &  (10)                  &          0      & (10)  \\
                                           &                                &                         &                          &          \\
$\sigma$ Ori         &  23\% \tablenotemark{b}       &      (11)              &             17          &  (12)        \\
                                           &                                &                        &                           &          \\                
 Lupus                 &    \ldots\tablenotemark{c}      &       (13)            &                 0        & (13)         \\
                                        &                                &                        &                           &          \\
$\rho$ Oph        &     35\% \tablenotemark{d}     &         (14,15)           &              23         & (16,17)        \\  
                                           &                                &                         &                           &        \\                          
\enddata
\tablenotetext{a}{Only isolated objects are listed here, thus leaving out TMR-1c in Taurus \citep{RIA11},
the candidate PMO companion to 2M J044144 \citep{TOD10}, and the candidate PMO companion
to Par-Lup3-4 \citep{COM11}.}
\tablenotetext{b}{for d$=$352pc; 14\% for d$=$440 pc}
\tablenotetext{c}{Although no substellar objects were found in the survey of (13), no definitive
value for this ratio can be derived, due to small number statistics.}
\tablenotetext{d}{Approximation using 110 candidate substellar members (14) and 316 members (15).}
\tablerefs{
(1) \citet{MUE03};
(2) \citet{LUH03}; 
(3) \citet{BUR09}; 
(4) \citet{SLE04};
(5) \citet{AND08};
(6) \citet{WEI09};
(7) \citet{ROB10};
(8) \citet{SCH09};
(9) \citet{MUZ11}; 
(10) \citet{LUH04} 
(11)\citet{LOD09};
(12) Fig. 2 of \citet{BIH09};
(13) \citet{COM11};
(14) \citet{ALV10};
(15) \citet{WIL08};
(16) \citet{MAB10};
(17) \citet{HAI10}
}
\end{deluxetable}

\clearpage

\begin{deluxetable}{rccrrrrrrrrrrrrrrrccc}
\rotate
\tiny
\tablecaption{Rho Ophiuchi Low Mass Candidate Members\tablenotemark{a}\label{table3}}
\tablewidth{0pt}
\tabletypesize{\tiny}
\tablehead{\colhead{ID}&\colhead{RA\tablenotemark{b}}&\colhead{Dec\tablenotemark{b}}&\colhead{$J$}&\colhead{${\sigma}J$}&\colhead{$H$}&\colhead{${\sigma}H$}&\colhead{$K$}&\colhead{${\sigma}K$}&\colhead{$I1$}&\colhead{${\sigma}I1$}&\colhead{$I2$}&\colhead{${\sigma}I2$}&\colhead{$I3$}&\colhead{${\sigma}I3$}&\colhead{$I4$}&\colhead{${\sigma}I4$}&\colhead{A$_{v}$\tablenotemark{c}}&\colhead{$Teff$\tablenotemark{d}}&\colhead{$Model$\tablenotemark{e}}&\colhead{IREX\tablenotemark{f}}}
\startdata
1&16:25:46.71&$-$24:18:13.6&17.92&0.07&16.48&0.04&15.31&0.04&0.351&0.034&0.254&0.023&0.418& 0.327&$-$1.150&1.540&3.3& 2099&B& E\\
2 & 16:25:46.74  & -24:19:15.9   & 19.98   & 0.19 & 18.27  & 0.08 & 16.84  & 0.07 & 0.059  & 0.018 & 0.042 & 0.013 & 0.102 & 0.157 & 0.163  & 0.288 & 5.6 & 1959 & B & E\\
3 & 16:25:46.80  & -24:19:56.4   & 19.29   & 0.13 & 17.80  & 0.06 & 16.83  & 0.07 & 0.110  & 0.022 & 0.061 & 0.027 & 0.328 & 0.247 & 0.850  & 0.462 & 5.6 & 2234 & B & E\\
4 & 16:25:46.84  & -24:17:58.5   & 18.15   & 0.08 & 16.25  & 0.03 & 14.96  & 0.03 & 0.569 & 0.040 & 0.414 & 0.030 & 0.121 & 0.345 & 0.658  & 1.070 & 8.7 & 2401 & B & E\\
5 & 16:25:46.92  & -24:20:32.8   & 19.46   & 0.16 & 18.01  & 0.07 & 17.00  & 0.07 & 0.078 & 0.024 & 0.049 & 0.014 & 0.078 & 0.164 & 0.172 & 0.324 & 4.8 & 2190 & B & E\\
\enddata
\tablenotetext{a}{Table 3 is published in its entirety in the electronic edition of the {\em Astrophysical Journal}. A portion is shown here for guidance regarding its form and content.}
\tablenotetext{b}{Coordinates listed are J2000. Units of right ascension are hours, minutes, and seconds, and units of declination are degrees, arcminutes, and arcseconds.}
\tablenotetext{c}{Extinction estimates were calculated by dereddening each source in the $JHK$ color-color diagram as discussed in the text.}
\tablenotetext{d}{Effective temperatures obtained from model fits as discussed in the text.}
\tablenotetext{e}{Model used for best fit. B = Blackbody, D = DUSTY, C=COND, N=NextGen. See text for details.}
\tablenotetext{f}{Infrared Excess (E) or Non-Excess (NE) source.}
\end{deluxetable}

\clearpage
\begin{deluxetable}    {cclcclcccccccccccccc}
\rotate 
\footnotesize
\tablecaption{Photometric Comparison with Recent Sensitive NIR Surveys of $\rho$ Oph}
\label{table4}
\tablewidth{0pt}
\tabletypesize{\scriptsize}
\tablehead{\multicolumn{3}{c}{RA(2000)} &  \multicolumn{3}{c}{Dec(2000)} &  \multicolumn{3}{c}{This work }  &
     \multicolumn{4}{c}{Alves de Oliveira et al.} & \multicolumn{4}{c}{Marsh et al.} & \multicolumn{3}{c}{Geers et al.} \\
     $h$& $m$& $s$& $^{\circ}$& $^{\prime}$& $^{\prime\prime}$& $J$& $H$& $K_s$& Source &$J$& $H$& $K_s$& Source & $J$& $H$& $K_s$ &
     Source& $J$& $K_s$\\
      & & & & & & & & &No. & & & &No. & & & &No. & & }
\startdata
16 & 25 & 57.71 & $-$24 & 23 & 17.56 & 17.89 & 14.92 & 13.04 &   8 & 17.68 & 14.84 &  13.11 &  &  &  &  & & &  \\
16 & 26 & 03.33 & $-$24 & 30 & 25.02 & 17.87 & 16.45 & 15.25 &   9 & 17.76 & 16.33 & 15.32 &  &  &  &  & & &  \\
16 & 26 & 04.56 & $-$24 & 17 & 51.32 & 15.51 & 13.45 & 12.06 &      &           &           &           &  &  &  &  & 12 & 15.79 & 12.19  \\
16 & 26 & 07.24 & $-$24 & 21 & 16.49 & 20.90 & 18.47 & 16.61 & 10 & 21.74 & 18.60 & 16.69 &  &  &  &  & &  &   \\  
16 & 26 & 07.92 & $-$24 & 17 & 22.89 & $\ge$21.00 & 18.90 & 17.16 & 11 & 22.26 & 19.25 & 17.31 &  &  &  &  &  &  &   \\  
16 & 26 & 11.69 & $-$24 & 24 & 30.98 & $\ge$21.00 & 18.17 & 16.13 & 13 & 21.20 & 18.45 & 16.13 &  &  &  &  &  &  &    \\  
16 & 26 & 13.16 & $-$24 & 19 & 09.71 & $\ge$21.00 & 18.26 & 16.38 & 14 & 22.17 & 18.77 & 16.40 &  &  &  &  &  &  &    \\
16 & 26 & 16.27 & $-$24 & 39 & 30.50 & 15.95 & 14.31 & 13.17 & 15 & 15.79 & 14.15 & 13.11 &  &  &  &  &  &  &   \\  
16 & 26 & 18.62 & $-$24 & 29 & 52.96 & 17.24 & 15.13 & 13.55 &      &           &           &           &  &  &  &  &1 & 17.36 & 13.61 \\
16 & 26 & 18.89 & $-$24 & 26 & 10.95 & 14.85 & 13.25 & 12.06 &    &       &       &       &  &  &  &  & 13 &  14.84 &  12.14 \\  
16 & 26 & 19.06 & $-$24 & 41 & 31.15 & 17.69 & 15.97 & 14.75 & 17 & 17.44 & 15.80 & 14.61 &  &  &  &  &  &  &   \\  
16 & 26 & 19.26 & $-$24 & 27 & 43.99 & 18.90 & 17.33 & 16.03 & 18 & 18.92 & 17.16 & 16.04 &  &  &  &  &  &  &   \\
16 & 26 & 21.46 & $-$24 & 26 & 00.76 & 12.43 & 11.53 & 10.73 &    &       &       &       &  &  &  &  & 14 &  12.57 &  10.92 \\  
16 & 26 & 22.25 & $-$24 & 37 & 08.27 & 18.98 & 16.88 & 15.64 & 19 & 19.15 & 17.10 & 15.86 &  &  &  &   &  &  &   \\
16 & 26 & 22.20 & $-$24 & 24 & 06.58 & 16.32 & 14.83 & 13.64 &  &  &  &  &  &  &  &  & 2 & 16.70  & 13.94   \\ 
16 & 26 & 23.84 & $-$24 & 18 & 28.30 & 16.06 & 14.64 & 13.40 &  &  &  &  &  &  &  &  & 16 & 16.07  & 13.58   \\ 
16 & 26 & 24.24 & $-$24 & 15 & 52.53 & 17.53\tablenotemark{1} & 15.48\tablenotemark{1} & 14.02\tablenotemark{1} & & & && &  & & & && \\   
16 & 26 & 24.32 & $-$24 & 15 & 48.07 & 18.60 & 15.90 & 13.75 &  21    & 18.78\tablenotemark{1} &15.85\tablenotemark{1} &13.83\tablenotemark{1}& &  & & & && \\       
16 & 26 & 25.04 & $-$24 & 41 & 33.51 & 17.23 & 15.69 & 14.61 & 22 & 17.03 & 15.52 & 14.06 &   &  &  &   &  &  &          \\
16 & 26 & 25.64 & $-$24 & 37 & 27.81 & 18.33  & 16.59 & 15.39 &  &  &  &  & 1045 & 18.201 & 16.544& 15.319 &  &  &   \\ 
16 & 26 & 25.98 & $-$24 & 33 & 13.87 & $\ge$21.00 & 20.75 & 18.57 & 23 & 22.20 & 19.39 & 17.11  &  &  &  &   &  &  &          \\
16 & 26 & 26.44 & $-$24 & 33 & 04.86 & $\ge$21.00 & 18.69 & 16.92 & 24 & 22.13 & 19.45 & 17.45  &  &  &  &   &  &  &       \\
16 & 26 & 27.76 & $-$24 & 26 & 41.60 & 14.04 & 12.91 & 11.96 &  &  &  &  &  &  &  &  & 17 & 14.26  & 12.09   \\ 
16 & 26 & 33.83 & $-$24 & 18 & 52.96 & $\ge$21.00 & 20.91 & 18.84 & 25 & 22.42 & 19.31 & 17.39  &  &  &  &   &  &  &  \\ 
16 & 26 & 34.00 & $-$24 & 35 & 55.88 & $\ge$21.00 & 19.44 & 17.47 & 26 &22.45 & 19.31 & 17.38 &  &  &  &  &  &  &   \\ 
16 & 26 & 35.37 & $-$24 & 30 & 11.15 & $\ge$21.00 & 18.01 & 15.93 & 27 & 22.32 & 18.35 & 15.94 &  &   &  &  &  &  &    \\ 
16 & 26 & 35.31 & $-$24 & 42 & 40.93 & 19.31 & 17.72 & 16.28 & 28 & 19.68 & 17.79 & 16.36 & &  &  &   &  &  &              \\
16 & 26 & 36.00 & $-$24 & 20 & 58.65 & 18.72 & 15.77 & 13.69 & 29 & 18.84 & 15.65 & 13.46 & &  &  &   &  &  &        \\
16 & 26 & 36.82 & $-$24 & 18 & 59.99 & 17.00 & 16.06 & 15.23 & 30 & 16.24 & 14.35 & 13.06 & &  &  &    & 3 &  17.01 & 14.55   \\   
16 & 26 & 37.84 & $-$24 & 39 & 03.20 & 14.85 & 13.73 & 12.74 & 31 & 14.65 & 13.45 & 12.66 & &  &  &    & 18 &  14.98 & 12.85   \\   
16 & 26 & 38.78 & $-$24 & 23 & 22.20 & 14.94 & 12.81 & 11.42 &    &      &       &       & &  &  &    & 2-3\tablenotemark{5} &  15.10 & saturated  \\  
16 & 26 & 39.61 & $-$24 & 18 & 02.90 & $\ge$21.00 & 18.67 & 17.18 & 32 &  22.55 & 19.15 & 17.11 & & & &  &  &   & \\   
16 & 26 & 39.68 & $-$24 & 22 & 07.51 & 18.44 & 17.04 & 15.93 & 33 & 18.16 & 16.74 & 15.68  &  & & &  &  &   &      \\  
16 & 26 & 39.92 & $-$24 & 22 & 32.43 & 16.19 & 14.66 & 13.43 & 34 & 15.95 & 14.53 &  & & &  &  &   & \\
16 & 26 & 40.03 & $-$24 & 28 & 07.37 & $\ge$21.00 & 18.32 & 16.58 & 35 &  21.45 & 18.56 & 16.52 &  & & &  &  &   &  \\
16 & 26 & 40.58 & $-$24 & 24 & 27.26 & 20.71 & 18.25 & 15.71 & 36 & 21.60 & 18.37 & 16.69 & &  & &  &  &   &  \\
16 & 26 & 40.85 & $-$24 & 30 & 50.62 & 17.44 & 14.89 & 13.12 & 37 & 17.32 & 14.77 & 13.18 &  & & &  &  &   &  \\ 
16 & 26 & 41.80 & $-$24 & 36 & 11.50 & $\ge$21.00 & 20.16 & 17.77 & 38 & 22.12 & 18.62 & 16.28 & & & &  &  &   &          \\  
16 & 26 & 41.83 & $-$24 & 23 & 43.62 & $\ge$21.00 & 18.62 & 17.04 & 39 & 22.59 & 19.32 & 17.07 & & &  &  &   &         \\
16 & 26 & 42.73 & $-$24 & 24 & 27.15 & 19.66 & 15.57 & 13.15 & 40 & 19.44 & 15.59 & 13.22 & & & &  &  &   &    \\
16 & 26 & 43.78 & $-$24 & 24 & 50.95 & $\ge$21.00 & 17.41 & 14.73 & 41 & 21.67 & 17.50 & 14.76 & & & &  &  &   &   \\                         
16 & 26 & 48.45 & $-$24 & 28 & 36.12 & 19.16 & 15.23 & 12.70 & 43 & 19.31 & 15.19 & 12.66 & & &  & &  &   &        \\ 
16 & 26 & 48.75 & $-$24 & 26 & 25.80 & 19.69 & 15.32 & 12.89 & 44 & 19.32 & 15.32 & 12.92 & & & & & &  &   \\
16 & 26 & 50.91 & $-$24 & 26 & 07.67 & $\ge$21.00 & 19.14 & 17.12 & 45 & 21.73 & 19.17 & 17.27 & & & & & &  &   \\
16 & 26 & 51.22 & $-$24 & 32 & 41.43 & 15.02 & 14.46 & 13.76 &    &      &       &       & & & & & 19 & 15.30 & 13.89   \\
16 & 26 & 51.91 & $-$24 & 30 & 38.62 & $\ge$21.00 & 16.32 & 13.33 & 46 & 21.30 & 16.43 & 13.45 & & & & & &  &        \\ 
16 & 26 & 52.70 & $-$24 & 24 & 52.85 & $\ge$21.00 & 18.30 & 15.78 & 48 & 21.76 & 18.24 & 15.80 & & & & & &  &         \\ 
16 & 26 & 53.43 & $-$24 & 32 & 35.67 & 20.88 & 16.65 & 13.29 & 49 & 21.83 & 16.69 & 13.30 & & & &  & &  & \\
16 & 26 & 54.33 & $-$24 & 24 & 38.59 & $\ge$21.00 & 17.16 & 14.02 & 50 & 21.71 & 17.25 & 14.04 & & & & & & & \\ 
16 & 26 & 54.74 & $-$24 & 27 & 02.40 & 17.85 & 14.76 & 12.69 & 51 & 17.92 & 14.91 & 12.87 & & & & & &    &         \\ 
16 & 26 & 55.34 & $-$24 & 21 & 17.10 & $\ge$21.00 & 18.37 & 17.23 &  &  &  &  & 4220 & 19.771 & 18.361 & 17.317 &  & &   \\ 
16 & 26 & 55.47 & $-$24 & 28 & 22.42 & 19.85 & 17.21 & 15.54 &  &  &  & & 1518 & 19.105 & 17.075 & 15.601 & & & \\
16 & 26 & 56.24 & $-$24 & 16 & 18.05 & 14.34 & 12.53 & 11.69 &  &  &  &  &108 & 14.373 & 12.424 & 11.615 & & & \\
16 & 26 & 56.25 & $-$24 & 21 & 30.90 & $\ge$21.00 & 18.75 & 17.34 &  &  &  &  & 4795 & 20.573 & 19.041 & 17.381 &  & &   \\ 
16 & 26 & 56.32 & $-$24 & 42 & 38.10 & 17.53  & 16.55 & 15.70 &  &  &  &  & 1254 & 17.753 & 16.633& 15.865 & 2-1\tablenotemark{5} & 17.68 & 15.53  \\ 
16 & 26 & 56.36 & $-$24 & 41 & 19.85 & 18.63 & 16.42 & 14.91 & 53 & 18.60 & 16.34 & 14.92 & 829 & 18.567 & 16.309 & 14.981 & 4 & 18.70 & 14.77 \\ 
16 & 26 & 56.87 & $-$24 & 28 & 36.33 & 18.20 & 14.72 & 12.75 &  &  &  &  & 233 & 18.256 & 14.690 & 12.800 &  &   &      \\
16 & 26 & 57.34 & $-$24 & 35 & 38.11 & 19.30 & 15.22 & 12.79 &  &  &  &  & 236 & 19.198 & 15.111 & 12.782 &  &   &      \\
16 & 26 & 57.37 & $-$24 & 42 & 18.70 & 18.86 & 17.15 & 15.99&  &  &  &  &     &        &        &        &  2-2\tablenotemark{5} & 18.94 & 15.73 \\
16 & 26 & 58.43 & $-$24 & 20 & 03.73 & $\ge$21.00 & 18.13 & 16.86 & 54 & 20.55 & 18.12 & 16.68 & & & & & &    &      \\   
16 & 26 & 58.35 & $-$24 & 21 & 30.28 & 16.09 & 13.10 & 11.61 & & & & & 115 & 16.299 & 13.178 & 11.515 & & & \\
16 & 26 & 58.67 & $-$24 & 24 & 55.47 & 20.25 & 17.25 & 14.91 & 55 & 20.22 & 17.15 & 14.84 & & & & & 5 & 19.64 & 15.30  \\
16 & 26 & 59.06 & $-$24 & 35 & 56.54 & 17.15 & 14.00 & 12.08 & & & &  &141 & 16.459 & 13.464 & 11.882 & 6 & 16.51 & 12.21  \\
16 & 26 & 59.94 & $-$24 & 24 & 21.62 & $\ge$21.00 & 19.38 & 18.39 &  &  &  & & 7614 & 20.740 & 19.286 &  & & & \\  
16 & 27 & 01.91 & $-$24 & 22 & 06.47 & 20.03 & 15.62 & 13.48 &  &  &  &  & 291 & 19.853 & 15.573 & 13.196 &  &   &      \\  
16 & 27 & 02.99 & $-$24 & 26 & 14.68 & $\ge$21.00 & 15.72 & 12.70 & & & & & 207 & \ldots & 15.673 & 12.559 & & & \\
16 & 27 & 03.57 & $-$24 & 20 & 05.11 & 17.32 & 15.13 & 13.97 & 56 & 17.13 & 15.01 & 13.72 & 311 & 17.067 & 14.738 & 14.437 & & & \\ 
16 & 27 & 04.09 & $-$24 & 28 & 30.23 & 16.43 & 13.03 & 10.92 &      & & & & 89 & 16.553 & 13.023 & 10.889 & & & \\
16 & 27 & 04.54 & $-$24 & 19 & 44.24 & $\ge$21.00 & 18.56 & 17.71 &  &  &  &  & 5598 & 20.239 & 18.642 & 17.607 &  & &   \\ 
16 & 27 & 04.56 & $-$24 & 27 & 15.21 & 16.98 & 13.47 & 11.47 & & & & & 103 & 16.464 & 13.076 & 11.185 & & & \\
16 & 27 & 05.64 & $-$24 & 40 & 12.85 & 20.94 & 16.72 & 14.01 &  &  &  &  & 439 & 21.353 & 16.645 & 13.911 & &  &\\ 
16 & 27 & 05.93 & $-$24 & 18 & 40.18 & 17.34 & 15.94 & 15.05 & 58 & 17.05 & 15.77 & 14.85 & 654 & 17.307 & 15.784 & 14.816 &  &  &   \\ 
16 & 27 & 05.97 & $-$24 & 28 & 36.73 & 16.86 & 14.58 & 13.09 &  &  &  & &      &        &       &        & 20 & 16.64 & 12.97 \\
16 & 27 & 05.98 & $-$24 & 16 & 14.15 & 18.76 & 17.21 & 16.19 &  &  &  & & 1344 & 18.889 & 17.069 & 16.032 & & & \\
16 & 27 & 06.52 & $-$24 & 18 & 32.18 &  20.19  & 18.35 & 17.63 &  &  &  & & 6412 & 19.965 & 18.600 &  & & & \\      
16 & 27 & 06.62 & $-$24 & 41 & 49.86 & 12.27 & 11.43 & 10.68 & & & & & 60 & 12.400 & 11.345 & 10.709 & & & \\
16 & 27 & 07.00 & $-$24 & 31 & 05.84 &  $\ge$21.00 & 18.50 & 15.94 &  &  &  & & 1558 & 21.587 & 18.758 & 15.909 & & & \\
16 & 27 & 07.68 & $-$24 & 34 & 03.04 & $\ge$21.00 & 18.63 & 16.75 & 59 & 22.33 & 19.00 & 16.58 & &  &  & & &  & \\
16 & 27 & 08.03 & $-$24 & 20 & 06.87 & 17.83 & 14.99 & 13.63 &  &  &  &  & 312 & 17.768 & 14.851 & 13.397 & & & \\ 
16 & 27 & 08.05 & $-$24 & 31 & 42.32 & $\ge$21.00 & 19.18 & 18.12 &  &  &  & & 7145 & 19.906 & 18.942 & 18.168 & & & \\  
16 & 27 & 08.14 & $-$24 & 41 & 18.94 & 20.49 & 18.40 & 17.13 &  &  &  &  & 3253 & 19.696 & 18.467 & 16.800  &  & &   \\ 
16 & 27 & 08.22 & $-$24 & 42 & 29.97 & 15.25 & 12.31 & 10.81 & & & & & 74 & 15.393 & 12.261 & 9.656\tablenotemark{2} & & & \\
16 & 27 & 08.44 & $-$24 & 16 & 19.61 & 20.44 & 18.30 & 17.17 &  &  &  &  & 4933 & 20.158 & 18.726 & 17.468 &  & &   \\ 
16 & 27 & 09.01 & $-$24 & 30 & 25.31 & $\ge$21.00 & 18.61 & 16.02 & 60 & 22.14 & 19.30 &  16.15& & & & & & &  \\ 
16 & 27 & 09.33 & $-$24 & 24 & 04.58 & 20.97 & 18.47 & 17.19 &  &  &  &  & 4788 & 19.401 & 18.166 & 17.591 &  & &   \\ 
16 & 27 & 09.36 & $-$24 & 32 & 15.24 &  $\ge$21.00  & 18.92 & 16.95 &  &  &  & & 2438 & 19.701 & 18.018 & 16.741 & & & \\
16 & 27 & 09.59 & $-$24 & 24 & 17.93 & $\ge$21.00 & 18.59 & 18.17 &  &  &  &  & 5710 & 19.606 & 18.248 & 18.566 &  & &   \\ 
16 & 27 & 09.80 & $-$24 & 34 & 41.27 & $\ge$21.00 & 20.45 & 18.15 & 61 & 22.44 & 18.84 & 16.51 & & & & & & &\\ 
16 & 27 & 10.05 & $-$24 & 29 & 13.55 & 16.47 & 15.16 & 14.05 & 62 & 16.55 & 15.14 & 14.23 &  &  & & & &   &     \\   
16 & 27 & 10.20 & $-$24 & 35 & 45.89 & $\ge$21.00 & 18.63 & 17.13 & 63 & 22.01 & 18.92 & 17.10 & & & & & & &  \\ 
16 & 27 & 10.33 & $-$24 & 33 & 22.32 & 17.32 & 13.92 & 11.92 &  &  &  &  & 147 & 17.401 & 13.854 & 11.910 &  &  &  \\
16 & 27 & 11.28 & $-$24 & 23 & 27.15 & $\ge$21.00 & 18.81 & 17.78 &  &  &  &  & 4863 & 19.503 & 18.568 & 17.460 &  & &   \\ 
16 & 27 & 11.60 & $-$24 & 23 & 21.80 & 20.31 & 18.16 & 17.19 &  &  &  &  & 4195 & 19.855 & 18.551 & 17.405 &  & &   \\ 
16 & 27 & 11.64 & $-$24 & 23 & 42.28 & 14.40 & 11.84 & 10.48 & & & & & 58 & 14.283 & 11.616 & 10.089 & & & \\
16 & 27 & 11.95 & $-$24 & 26 & 46.61 &  $\ge$21.00  & 18.81 & 17.99 &  &  &  & & 5820 & 20.136 & 18.657 & 17.925 & & & \\
16 & 27 & 12.71 & $-$24 & 32 & 0.00 &  $\ge$21.00 & 19.26 & 18.20 &  &  &  & & 14250 & 19.573 & 18.885 & 18.300 & & & \\  
16 & 27 & 13.01 & $-$24 & 31 & 59.99 & $\ge$21.00 & 18.67 & 16.92 & 64 & 22.13 & 19.28 & 17.05 & 2978 & 19.474 & 18.380 & 16.840  &  & &   \\ 
16 & 27 & 13.17 & $-$24 & 23 & 47.60 & $\ge$21.00 & 17.67 & 15.53 & 65 & 20.88 & 17.60 & 15.67 & & & & & & & \\ 
16 & 27 & 13.55 & $-$24 & 34 & 14.43 & $\ge$21.00 & 19.57 & 18.21 &   & & & & 9002 & 20.211 & 19.772 & 18.713 & & & \\
16 & 27 & 13.99 & $-$24 & 32 & 06.19 & $\ge$21.00 & 18.94 & 17.62 &  &  &  &  & 5076 & 20.054 & 19.112 & 17.741 &  & &   \\ 
16 & 27 & 14.08 & $-$24 & 22 & 50.59 & 20.98  & 18.26 & 16.58 &  &  &  & & 2956 & 20.144 & 18.405 & 16.759 & & & \\
16 & 27 & 14.31 & $-$24 & 31 & 31.85 & 18.36 & 16.38 & 15.02 & 66 & 18.42 & 16.53 & 15.30 & & & & & & & \\
16 & 27 & 15.73 & $-$24 & 38 & 43.68 & 14.17 & 11.62 & 10.17 & 67 & 18.76 & 15.22 & 12.95 & & & & & & &  \\
16 & 27 & 15.83 & $-$24 & 25 & 13.93 & 20.80 & 16.02 & 13.22 & 68 & 20.23 & 15.90 & 13.26 & 313 & 20.289 & 15.988 & 13.283 &  &     &      \\
16 & 27 & 15.83 & $-$24 & 34 & 06.74 &  $\ge$21.00  & 19.07 & 17.49 &  &  &  & & 5771 & 20.936 & 19.247 & 17.686 & & & \\
16 & 27 & 15.90 & $-$24 & 22 & 53.23 & $\ge$21.00 & 19.73 & 17.94 &  &  &  & & 7906 & 21.006 & 19.460 & 18.198  & & & \\  
16 & 27 & 17.38 & $-$24 & 32 & 06.97 &  $\ge$21.00 & 17.84 & 15.73 &  &  &  & & 1604 & 19.552 & 17.594 & 15.848 & & & \\
16 & 27 & 17.40 & $-$24 & 22 & 28.27 & 20.14 & 15.90 & 13.56 &  &  &  &  & 341 & 19.672 & 15.605 & 13.430 & &  &\\ 
16 & 27 & 18.17 & $-$24 & 25 & 55.52 & $\ge$21.00 & 18.45 & 16.93 &  &  &  &  & 3872 & 20.056 & 18.355 &  &  & &   \\ 
16 & 27 & 18.33 & $-$24 & 24 & 25.75 & 16.87 & 13.33 & 11.28 & & & & & 107 & 16.907 & 13.229 & 11.271 & & & \\
16 & 27 & 19.37 & $-$24 & 20 & 49.25 & 20.89 & 18.30 & 16.24 & 71 & 21.73 & 18.35 & 16.17 & & & & & & & \\ 
16 & 27 & 19.44 & $-$24 & 26 & 00.82 &  $\ge$21.00 & 18.34 & 16.72 & 70 &  21.71 & 18.71 & 16.79 & & &  & & &    &     \\
16 & 27 & 19.54 & $-$24 & 26 & 21.28 & $\ge$21.00 & 18.63 & 18.04 &  &  &  &  & 5454 & 20.136 & 18.466 & 17.857  &  & &   \\ 
16 & 27 & 19.79 & $-$24 & 26 & 35.50 &  $\ge$21.00 & 19.46 & 18.50 &  &  &  & & 14819 & 20.280 & 19.272 & 18.697 & & & \\  
16 & 26 & 21.07 & $-$24 & 28 & 28.55 & $\ge$21.00 & 18.79 & 17.30 &  &  &  &  & 4823 & 20.559 & 19.048 & 17.432 &  & &   \\ 
16 & 27 & 21.21 & $-$24 & 37 & 53.59 &  $\ge$21.00 & 18.87 & 16.84 & 73 & 22.83 & 19.47 & 17.36 & & & & & &  &     \\
16 & 27 & 21.55 & $-$24 & 21 & 50.85 & 15.59 & 12.63 & 10.98 & & & & & 90 & 15.645 & 12.588 & 11.016 & & & \\
16 & 27 & 21.66 & $-$24 & 32 & 17.93 &  20.97 & 18.30 & 16.40 &  &  &  & & 2403 & 18.968 & 17.650 & 16.421 & & & \\
16 & 27 & 21.99 & $-$24 & 29 & 38.29 & 19.77 & 15.41 & 12.90 &  &  &  &  & 238 &             & 15.424 & 12.838 &  &   &      \\
16 & 27 & 22.46 & $-$24 & 38 & 37.44 &  $\ge$21.00 & 18.63 & 16.96 & 74 & 22.20 & 19.19 & 16.92 & & & & & &  &       \\ 
16 & 27 & 22.90 & $-$24 & 18 & 26.38 &  $\ge$21.00  & 19.26 & 18.03 &  &  &  & & 5757 & 20.668 & 19.698 & 17.982 & & & \\
16 & 27 & 22.97 & $-$24 & 22 & 37.01 & $\ge$21.00 & 18.79 & 17.20 &  &  &  &  & 3809 & 20.587 & 18.893 & 17.066  &  & &   \\ 
16 & 27 & 23.58 & $-$24 & 30 & 46.55 & $\ge$21.00 & 19.13 & 17.26 &  &  &  &  & 4264 & 19.205 & 18.177 & 17.286 &  & &   \\ 
16 & 27 & 23.59 & $-$24 & 34 & 44.30 & $\ge$21.00 & 19.38 & 18.03 &  &  &  & & 6745 & 19.838 & 19.049 & 18.329 & & & \\  
16 & 27 & 24.17 & $-$24 & 25 & 10.86 & $\ge$21.00 & 18.16 & 15.81 & 75 & 22.32 & 18.73 & 15.85 & & & & & & & \\
16 & 27 & 24.29 & $-$24 & 20 & 44.72 & $\ge$21.00 & 18.69 & 17.05 &  &  &  &  & 5084 & 20.888 & 19.267 &  &  & &   \\ 
16 & 27 & 24.37 & $-$24 & 41 & 48.29 & 18.91 & 15.04 & 12.70 & 76 & 18.78 & 15.01 & 12.72 & 222 & 18.556 & 15.001 & 12.692 &  &   &      \\
16 & 27 & 24.60 & $-$24 & 28  & 49.30 & $\ge$21.00 & $\ge$21.00 & 18.50 &  & & & & 4077 & 19.920 & 18.673 & & & & \\
16 & 27 & 24.64 & $-$24 & 34 & 22.04 &  $\ge$21.00  & 18.15 & 16.45 &  &  &  & & 2448 & 19.868 & 17.934 & 16.415 & & & \\
16 & 27 & 24.67 & $-$24 & 29 & 34.12 & 19.37 & 14.88 & 12.48 & & & & & 195 & 19.268 & 14.836 & 12.474 & & & \\
16 & 27 & 25.42 & $-$24 & 25 & 37.51 & $\ge$21.00 & 18.76 & 18.15 &  &  &  &  & 4450\tablenotemark{3} & 19.573 & 18.335 & 17.709 &  & &   \\ 
16 & 27 & 25.62 & $-$24 & 35 & 05.79 & $\ge$21.00 & 19.22 & 18.41 &  &  &  & & 7704 & 19.653 & 19.054 & 18.642  & & & \\  
16 & 27 & 25.64 & $-$24 & 37 & 27.81 & 18.33 & 16.59 & 15.23 & 77 & 18.23 & 16.54 & 15.36 & & & & & & & \\
16 & 27 & 25.75 & $-$24 & 29 & 53.59 & $\ge$21.00 & 18.34 & 17.43 &  &  &  &  & 4114 & 19.204 & 18.119 & 17.253 &  & &   \\ 
16 & 27 & 25.97 & $-$24 & 28 & 56.75 & $\ge$21.00 & 19.52 & 17.83 &  &  &  &  & 2993 & 20.532 & 18.752 & 17.176  &  & &   \\ 
16 & 27 & 26.18 & $-$24 & 19 & 23.03 & 16.42 & 14.32 & 12.83 & 78 & 16.35 & 14.36 & 13.04 & 239 & 16.606 & 14.343 & 12.977 & 21 & 16.40  & 12.93     \\  
16 & 27 & 26.57 & $-$24 & 25 & 54.39 & 12.79 & 12.31 & 11.58 & & & & & 93 & 13.033 & 12.373 & 11.909 & 22 & 13.00 & 11.84  \\
16 & 27 & 26.62 & $-$24 & 40 & 45.13 & 18.61 & 14.94 & 12.60 & 79 & 18.49 & 14.89 & 12.54 & & & & &  &   &   \\     
16 & 27 & 27.09 & $-$24 & 32 & 16.95 & 18.20 & 14.49 & 12.37 & & & & & 183 & 18.354 & 14.465 & 12.384 & & & \\
16 & 27 & 27.68 & $-$24 & 38 & 26.95 & $\ge$21.00 & 18.50 & 16.91 & 80 & 21.94 & 18.94 & 16.95 & & & & & & & \\
16 & 27 & 28.18 & $-$24 & 31 & 42.23 & 20.85 & 16.14 & 13.80 &  &  &  &  & 408 & 20.695 & 16.032 & 13.760 & &  &\\ 
16 & 27 & 28.71 & $-$24 & 17 & 06.52 & 18.64 & 14.86 & 12.80 &  &  &  &  & 247 & 18.784 & 14.866 & 12.892 &  &   &      \\  
16 & 27 & 29.30 & $-$24 & 34 & 07.97 & 17.98 & 14.69 & 12.82 &  &  &  &  & 232 & 17.831 & 14.644 & 12.828 &  &   &      \\
16 & 27 & 29.46 & $-$24 & 39 & 15.95 & 16.37 & 12.52 & 9.95\tablenotemark{2} & & & & & 62 & 17.088 & 12.966 & 10.162 & & & \\
16 & 27 & 29.52 & $-$24 & 19 & 44.80 & $\ge$21.00 & 19.70 & 18.63 &  &  &  & & 9096 & 20.968 & 19.852 & 18.424  & & & \\  
16 & 27 & 29.68 & $-$24 & 29 & 24.75 & $\ge$21.00  & 16.99 & 14.47 &  &  &  &  & 681 & 19.824 & 16.859 & 14.504 &  &  &   \\ 
16 & 27 & 30.56 & $-$24 & 38 & 26.43 & $\ge$21.00 & 19.12 & 18.42 &  &  &  & & 6740 & 19.418 & 18.886 & 17.647 & & & \\  
16 & 27 & 30.62 & $-$24 & 32 & 34.41 & 12.94 & 12.55 & 12.08 & & & & & 101 & 13.022 & 12.453 & 12.213 & & & \\
16 & 27 & 30.96 & $-$24 & 20 & 01.74 & $\ge$21.00 & 18.12 & 17.06 &  &  &  &  & 3754 & 20.806 & 19.180 & 17.059  &  & &   \\ 
16 & 27 & 31.07 & $-$24 & 34 & 02.82 & 13.35 & 11.40 & 10.31 & & & & & 59 & 13.431 & 11.327 & 10.344 & & & \\
16 & 27 & 31.77 & $-$24 & 31 & 48.20 & $\ge$21.00 & 18.22 & 16.52 & 83 & 21.92 & 18.47 &  16.44 & & & & & & &\\
16 & 27 & 32.19 & $-$24 & 29 & 42.79 & 18.67 & 15.08 & 12.95 & 84 & 18.35 & 15.04 & 13.05 & 334 & & 15.088 &12.995 & 7 & 18.47 & 13.41\\ 
16 & 27 & 32.51 & $-$24 & 16 & 04.20 &  $\ge$21.00 & 18.14 & 16.39 & 85 & 22.42 & 18.81 & 16.57 & &  & & & & &    \\ 
16 & 27 & 32.53 & $-$24 & 39 & 46.07 & 20.87  & 18.11 & 16.35 &  &  &  & & 2797 & 19.716 & 17.647 & 16.255 & & & \\
16 & 27 & 32.70 & $-$24 & 33 & 23.63 & 16.20 & 12.87 & 11.03 &  &  &  &  & 132 & 16.126 & 12.685 & 10.848 & & & \\
16 & 27 & 32.70 & $-$24 & 22 & 46.48 &  20.81  & 18.17 & 17.20 &  &  &  & & 6419 & 19.441 & 18.436 & 17.425 & & & \\  
16 & 27 & 32.73 & $-$24 & 32 & 41.83 & 19.00 & 15.21 & 13.04 &  &  &  &  & 491 & 19.231 & 15.795 & 13.377 & &  &\\ 
16 & 27 & 32.96 & $-$24 & 28 & 11.15 & 20.87  & 17.45 & 15.02 &  &  &  &  & 1307 & 19.781 & 17.189 & 14.922 &  &  &   \\ 
16 & 27 & 33.55 & $-$24 & 22 & 49.12 & 20.37 & 18.25 & 17.12 &  &  &  & & 9558 & 19.768 & 19.332 & 17.708 & & & \\  
16 & 27 & 33.67 & $-$24 & 30 & 50.99 & $\ge$21.00 & 18.31 & 17.24 &  &  &  &  & 5003 & 19.691 & 18.954 & 16.895 &  & &   \\ 
16 & 27 & 33.81 & $-$24 & 22 & 34.33 & 17.84  & 16.76 & 15.69 &  &  &  & & 2870 & 17.725 & 16.491 & 15.672 & & & \\
16 & 27 & 34.14 & $-$24 & 33 & 08.37 &  $\ge$21.00 & 18.50 & 17.14 & 87 & 22.22 & 18.96 & 17.03 & &  & & &  & &  \\ 
16 & 27 & 35.32 & $-$24 & 39 & 57.61 &  19.50 & 16.59 & 15.08 &  &  &  & & 2391 & 17.625 & 16.288 & 14.858 & & & \\
16 & 27 & 37.20 & $-$24 & 34 & 34.12 &  $\ge$21.00 & 18.18 & 16.02 & 92 & 21.92 & 18.21 & 15.97 & & & & & & & \\
16 & 27 & 37.24 & $-$24 & 25 & 26.43 &  $\ge$21.00 & 18.23 & 16.24 & 93 & 22.31 & 18.81 & 16.18 & & & & & &  &\\
16 & 27 & 37.40 & $-$24 & 17 & 54.78 & 14.08 & 12.82 & 11.80 & & & & &    &        &       &       & 23 & 14.15 & 11.95  \\
16 & 27 & 38.98 & $-$24 & 40 & 20.53 &  16.54 & 14.20 & 12.47 & 94 & 16.48 & 14.12 & 12.56  & & & & & 8, 24\tablenotemark{4} & 16.54 & 12.29 \\
16 & 27 & 40.13 & $-$24 & 26 & 36.60 & $\ge$21.00 & 17.30 & 14.15 & 95 & 21.90 & 17.39 & 14.13 & & & & & &  &\\
16 & 27 & 40.95 & $-$24 & 28 & 59.55 & 14.74 & 13.88 & 13.07 & 96 & 14.60 & 13.76 & 13.19 & & & & & 25 & 14.66 & 13.10  \\ 
16 & 27 & 41.84 & $-$24 & 42 & 34.99 & 20.38 & 17.47 & 15.18 &   &       &       &      & & & & & 9 & 21.32 & 14.97  \\ 
16 & 27 & 46.39 & $-$24 & 31 & 41.02 & 13.78 & 12.28 & 11.23 &   &       &       &      & & & & & 26 & 13.83 & 11.32  \\ 
16 & 27 & 56.76 & $-$24 & 28 & 16.74 &  $\ge$21.00  & 19.47 & 18.22 &  &  &  & & 6249 & 20.940 & 19.558 & 18.104 & & & \\                   
\enddata
\tablenotetext{1}{Source 21 of Table 4 of Alves de Oliveira et al. is a 4.5$^{\prime\prime}$ binary in our images--their coordinates fall between those of the two components,
whereas their photometry agrees with our photometry for the fainter component. We list the coordinates and photometry for the primary for completeness.}
\tablenotetext{2}{Saturated in our data.}
\tablenotetext{3}{This is the planetary mass object confirmed spectroscopically by \citet{MAB10}. See text for discussion.}
\tablenotetext{4}{Sources 8 and 24 as listed in Geers et al.'s Table 1 have identical entries for their coordinates and photometry,
with the exception of the $K_s$ photometry which is listed at 13.36 for Source 8 and as 12.29 for Source 24.  We have used the value
listed for Source 24, as it agrees better with other determinations.}
\tablenotetext{5}{Source No. from Table 2 of Geers et al. with spectroscopic follow-up.}
\end{deluxetable} 

\clearpage

\end{document}